\documentclass[twocolumn,english,showpacs,nofootinbib,superscriptaddress]{revtex4-1}
\usepackage[T1]{fontenc}
\usepackage[latin9]{inputenc}
\usepackage{geometry}
\usepackage{xcolor}
\usepackage{babel}
\usepackage{units}
\usepackage{amsmath}
\usepackage{amssymb}
\usepackage{bbold}
\usepackage{stmaryrd}
\usepackage{graphicx}
\usepackage{wasysym}
\usepackage{url}

\usepackage[bookmarks=false]{hyperref}

\makeatletter

\IfFileExists{lmodern.sty}{\usepackage{lmodern}}{}
\usepackage{babel}

\usepackage{hyperref}

\makeatother

\begin{document}
\title{Programmable N-body interactions with trapped ions}
\date{\today}

\author{Or Katz}
\email{Corresponding author: or.katz@duke.edu}
\address{Duke Quantum Center, Duke University, Durham, NC 27701}
\address{Department of Electrical and Computer Engineering, Duke University, Durham, NC 27708}
\address{Department of Physics, Duke University, Durham, NC 27708}
\author{Marko Cetina}
\address{Duke Quantum Center, Duke University, Durham, NC 27701}
\address{Department of Physics, Duke University, Durham, NC 27708}

\author{Christopher Monroe}
\address{Duke Quantum Center, Duke University, Durham, NC 27701}
\address{Department of Electrical and Computer Engineering, Duke University, Durham, NC 27708}
\address{Department of Physics, Duke University, Durham, NC 27708}
\address{IonQ, Inc., College Park, MD  20740}

\begin{abstract}
Trapped atomic ion qubits or effective spins are a powerful quantum platform for quantum computation and simulation, featuring densely connected and efficiently programmable interactions between the spins.
While native interactions between trapped ion spins are typically pairwise, many quantum algorithms and quantum spin models naturally feature couplings between triplets, quartets or higher orders of spins. Here we formulate and analyze a mechanism that extends the standard M\o{}lmer-S\o{}rensen pairwise entangling gate and generates a controllable and programmable coupling between $N$ spins of trapped ions. We show that spin-dependent optical forces applied at twice the motional frequency generate a coordinate-transformation of the collective ion motion in phase-space, rendering displacement forces that are nonlinear in the spin operators. We formulate a simple framework that enables a systematic and faithful construction of high-order spin Hamiltonians and gates, including the effect of multiple modes of motion, and characterize the performance of such operations under realistic conditions.

\end{abstract}
\maketitle
\section{Introduction}
Ions in a linear Paul trap are a salient platform for simulation of quantum spin dynamics \cite{Monroe2021} and for computation of problems that are classically hard \cite{QEDC}. Internal electronic energy levels of individual ions can be used as qubits or effective spins that can be efficiently prepared, controlled and measured with high isolation from the environment. When trapped ions are laser-cooled and ordered into long chains, their Coulomb interaction gives rise to collective modes of motion between the ions. With the addition of optical \cite{CiracZoller1995} or near-field microwave \cite{Mintert2003} driving fields, the resultant force can depend upon the quantum spin state of the ions, thus generating spin-spin entanglement and allowing for control over their many-body quantum state.

The most prominent configuration for such entangling operations uses bichromatic optical fields, which exert spin-dependent forces and result in the accumulation of a geometric spin-dependent phase \cite{sorensen2000entanglement,Milburn2000,Solano1999}. This mechanism forms the basis for two-qubit M\o{}lmer S\o{}rensen (MS) gates widely used in trapped-ion quantum computers, as well as effective Ising couplings in trapped-ion based quantum simulators \cite{Monroe2021,porras2004effective,blumel2021power,shapira2018robust,leung2018robust,webb2018resilient,wang2022ultra,shapira2022robust}. This engineered Ising coupling features dense or even full connectivity between pairs of ions, owing to their collective vibrations in a chain, but it is limited to two-body interactions. 

Most quantum circuits and many spin models call for higher-order interactions. Examples including the simulation of molecular orbitals in quantum chemistry \cite{seeley2012bravyi,o2016scalable,nam2020ground,aspuru2005simulated,hempel2018quantum}, quantum simulations of lattice gauge theories \cite{banuls2020simulating,ciavarella2021trailhead,hauke2013quantum}, stabilizer operators in quantum error correction codes \cite{paetznick2013universal,kitaev2003fault}, spin models \cite{pachos2004three,muller2011simulating,motrunich2005variational,andrade2022engineering} and generic quantum algorithms \cite{vedral1996quantum, grover1996fast,Wang2001,PhysRevLett.102.040501,Espinoza2021,figgatt2017complete,marvian2022restrictions}. While sequential or parallel application of universal one- and two-body gate sets can generate arbitrary entangled many-body states, such constructions can carry overhead in the number of entangling operations or Trotterization steps \cite{Lloyd1996} and thereby be limited in the face of decoherence.

Recently, we proposed a mechanism to realize a native $N$-body interaction between trapped ion spins by squeezing a single vibrational mode of motion in a state-dependent manner \cite{Katz2022Nbody}. We considered optical spin-dependent forces that are applied synchronously at twice the motional frequency of a particular vibrational mode of motion, generating a family of $N$-body entangling interactions and gates that can be realized in a single step. In this paper, we extend that study by fully considering the coupling to multiple motional modes in the trapped ion crystal. While the conventional MS-type interaction can be straightforwardly extended to off-resonant forces and multi-mode operation \cite{Sorensen1999, Debnath2016,zhu2021interactive,stricker2022towards,schwerdt2022comparing,seetharam2021digital,erhard2021entangling}, the nonlinear nature of squeezing forces renders the vibratory and spin evolution nontrivial owing to the quadratic dependence of the phonon operators in the interaction Hamiltonian. The treatment of multimode and off-resonant squeezing operations is also important in practice, as the parametric forces driven at twice the motional mode frequencies are generally accompanied by nearby off-resonant forces that can play an important role in the dynamics \cite{Katz2022Nbody}. 

Here we analyze the application of time-dependent squeezing acting simultaneously on multiple motional modes of an trapped ion chain. We formulate and characterize the evolution of the ions spin and motional states, revealing a large toolbox of effective spin Hamiltonians and quantum gates. We identify a particular protocol for multimode squeezing and displacement forces to demonstrate particular applications, including the construction of the $N$-body stabilizer operator composed of a product of $N$ spin operators, as well as extensions of the $N$-bit Toffoli gate using multiple modes. Finally we outline and demonstrate new avenues to program and simulate Hamiltonians composed of multiple high-order terms in a single step. 

This paper is organized as follows. In Sec.~\ref{sec:formalism} we describe
the time-dependent interaction Hamiltonian coupling the ions' spins and motion. The resulting time evolution is composed of spin-dependent motional squeezing and displacements that are controlled by the optical fields. In Sec.~\ref{sec:Spin-Motion Evolution} we find that this evolution, in the Heisenberg picture, is described by a spin-dependant linear coordinate transformation in phase space. In Sec.~\ref{sec:loop_protocol} we use this linear transformation to construct a family of gates that act on the spins to generate N-body interactions that are robust to thermal motion of ions. In Sec.~\ref{sec:applications} we present two numerical examples of gates that entangle four spins in a chain of eleven ions and show that nearby off-resonant motional modes can be controlled via pulse-shaping. Finally, in Sec.~\ref{sec:discussion} we discuss the practical application of these gates to current trapped ions systems and their prospect in other quantum hardware.

\section{Interaction Hamiltonian}\label{sec:formalism}
\begin{figure}[t]
\begin{centering}
\includegraphics[width=7.5cm]{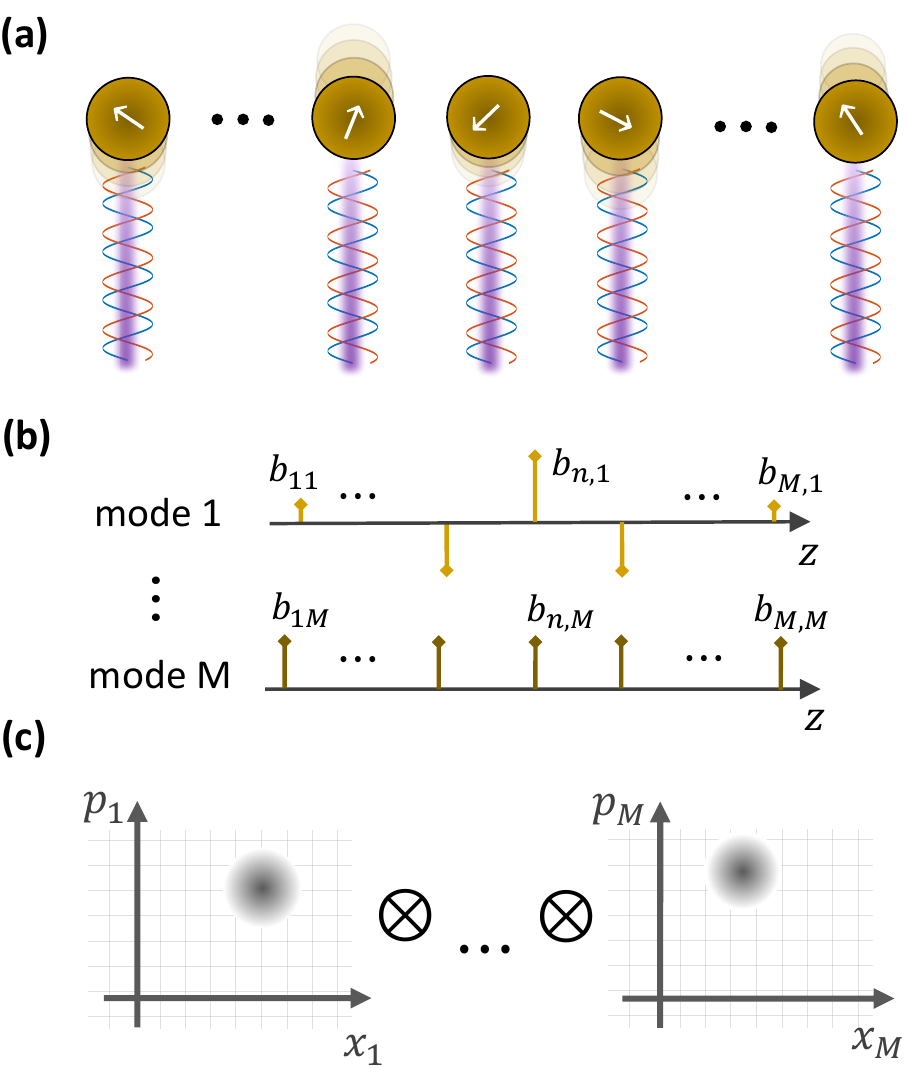}
\par\end{centering}
\centering{}\caption{\textbf{Trapped ions system.} (a) A crystal of $M$ ion spins trapped by external forces and addressed with an array of bi-chromatic optical fields. The beams apply state-dependent forces that couple the spin state of the ions with their motion. (b) Motion along the beam axis is composed of $M$ collective vibrational modes of the crystal. The matrix element $b_{nk}$ denotes the participation of the $n^{\textrm{th}}$ ion in the $k^{\textrm{th}}$ vibrational mode. (c) The vibrations of the crystal can be pictorially represented by $M$ phase-space diagrams with coordinates $\hat{x}_k$ and $\hat{p}_k$ for $1\leq k \leq M$. The coordinates are unitless (scaled to twice the zero-point position and momentum spreads) and described in the interaction frame, which rotates at the vibrational frequency $\omega_k$ of the $k^{\textrm{th}}$ mode, such that the motional state in phase space is stationary unless optical forces are applied. The shaded area represents an arbitrary motional state of the crystal in each phonon mode.  \label{fig:ion_trap_general}}
\end{figure}

We consider a linear chain of $M$ trapped atomic ions, each storing a spin-1/2 system, addressed by laser beams, as shown in Fig.~\ref{fig:ion_trap_general}. We assume that $M$ motional modes of the ions are aligned with the spatial direction of the applied optical forces. These modes are described by their frequencies $\omega_k$ and displacement eigenvectors $b_{nk}$ which describe the motional amplitude of the $n$th ion in the $k$th motional mode normalized such that $\sum_i b_{ij}b_{ik} = \delta_{jk}$ and $\sum_j b_{mj}b_{nj} = \delta_{mn}$. The phonon modes are characterized by the bosonic annihilation and creation operators $\hat{a}_k$ and $\hat{a}^\dagger_k$ of mode $k$, with $[\hat{a}_k,\hat{a}^\dagger_k]=1$. 

The applied optical fields couple the ion spins to their motion via the interaction Hamiltonian \cite{Leibfried2003}
\begin{equation}
H_{I}=\frac{\hbar}{2}\sum_{n=1}^M\tilde{\Omega}_{n}(t)\sigma_+^{(n)}\prod_{k=1}^Me^{i\eta_{nk}\left(\hat{a}_ke^{-i\omega_k t}+\hat{a}^\dagger_ke^{i\omega_k t}\right)} +\textrm{h.c.}\label{eq:interaction_hamiltonian_H_I},
\end{equation}
where the exponential term describes modulation of the optical phase in the oscillating ions' reference frame. Here, $\tilde{\Omega}_{n}(t)$ is the driving Rabi frequency for spin $n$ in a frame rotating at the frequency of the $n^{\textrm{th}}$ spin and $\sigma_\pm^{(n)}$ are the raising/lowering spin operators. The Lamb-Dicke parameters $\eta_{nk}=\delta K x^0_{m}b_{nk}$ describe the coupling between spin $n$ and mode $k$, where $\delta \textrm{k}$ is the effective wavenumber of the radiation field driving the sidebands and $x^0_{k}=\sqrt{\hbar/2\mathcal{M}\omega_k}$ is the zero-point spread in position of the $k$th phonon mode, taking $\mathcal{M}$ as the mass of a single ion \cite{Leibfried2003}. We assume that the radial motion along the optical beam is confined within the Lamb-Dicke regime where $|\eta_{nk}\langle \hat{a}_k^{\dagger}+\hat{a}_k \rangle| \ll 1$ for all ions and modes. 
While we have assumed the above spin-motion coupling originates from either a direct optical transition or a twin-beam optical Raman process between spin states \cite{Leibfried2003}, the framework here can also be applied to a microwave drive with field gradients \cite{Mintert2003,srinivas2021high,harty2016high,srinivas2019trapped}.

In phase space, $H_I$ acts on the motional state of the ions as shown in Fig.~\ref{fig:ion_trap_general}c. The phase-space coordinates of mode $k$ are defined by the unitless quadrature position and momentum operators $\hat{x}_k=(\hat{a}_k+\hat{a}^{\dagger}_k)/2$ and $\hat{p}_k=i(\hat{a}^{\dagger}_k-\hat{a}_k)/2$, which have been scaled by $2x^0_k$ and $2p^0_{k}=\sqrt{2\hbar \mathcal{M}\omega_k}$, respectively. These operators satisfy $[\hat{x}_k,\hat{p}_j]=i\delta_{jk}/2$. 

\begin{figure}[t]
\begin{centering}
\includegraphics[width=8.6cm]{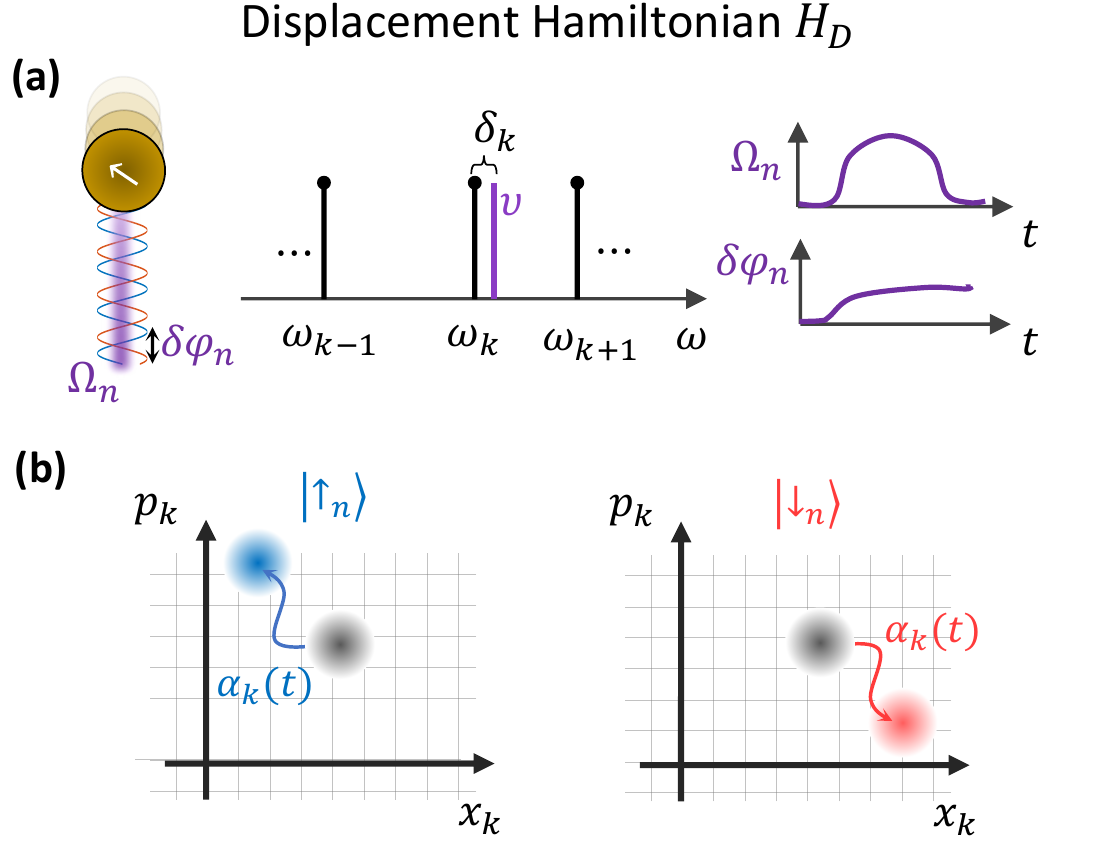}
\par\end{centering}
\centering{}\caption{\textbf{State-dependent displacement of motion.} \textbf{(a)} Tuning the frequency $\nu$ of the drive on the $n^{\textrm{th}}$ ion near the resonance frequency of the $k^{\textrm{th}}$ vibrational mode (upper and lower sidebands) couples the spin predominantly with the collective motion of that mode through the displacement Hamiltonian $H_D$ in Eq.~(\ref{eq:displacement_hamiltonian_H_D}). (b) In phase space, the evolution is represented by displacement of the collective ion motion by amount $\alpha_{k}$. The direction of displacement depends on the spin state and is inverted if the $n^{\textrm{th}}$ spin is flipped. The trajectory $\alpha_{k}(t)$ can be temporally engineered via modulation of the field amplitude $\Omega_n(t)$ and the relative phase $\delta \varphi_n$ between the upper and lower sideband tones (c.f.~Eq.~\ref{eq:displacemt_control}). The instantaneous amplitude of motion is controlled by the former whereas the instantaneous orientation of motion in phase space is controlled by the latter.
\label{fig:displacement_and_phase}}
\end{figure}

In this work, we focus on the symmetric driving of the the blue and red motional sidebands in Eq.~(\ref{eq:interaction_hamiltonian_H_I}). These bichromatic electromagnetic fields $\tilde{\Omega}_{n}=\Omega_{n}(t)\bigl[e^{-i(\nu t+\phi_{n+})}+e^{i(\nu t-\phi_{n-})}\bigr]$ are applied with frequencies $\pm\nu$ from the spin resonance carrier, with phases $\phi_{n\pm}$ and common amplitude $\Omega_{n}(t)$. 

Tuning near the first motional sidebands with $\delta_k \equiv \nu-\omega_k$ and $|\delta_k| \ll \omega_k$ generates the interaction Hamiltonian (under the rotating wave approximation where~$\Omega_n\ll\omega_k$) \cite{Monroe2021}
\begin{equation}
H_{D}=\frac{\hbar}{2}\sum_{m,n}\eta_{nm}\Omega_{n}(t)e^{i(\delta_m t+\delta\varphi_n)}{\sigma}_{\varphi_n}^{(n)}\hat{a}_{m} +\textrm{h.c.}.\label{eq:displacement_hamiltonian_H_D}
\end{equation} 
Here $\delta\varphi_n=(\phi_{n+}-\phi_{n-})/2$ is the relative phase between the two tones and $\bar{\varphi}_n=(\phi_{n+}+\phi_{n-}-\pi)/2$ is the common phase that determines the orientation of the spin operator on the Bloch sphere $\sigma_{\bar{\varphi_n}}^{(n)}=\cos\bar{\varphi}_n\sigma_{x}^{(n)}+\sin\bar{\varphi}_n\sigma_{y}^{(n)}$.
The Hamiltonian $H_D$ acts to displace the position and momentum of the phonon mode $k$ in a spin-dependent manner by $\mathrm{Re}(\alpha_k)$ and $\mathrm{Im}(\alpha_k$) respectively, where \begin{equation}\label{eq:displacemt_control} \alpha_k(t)=\tfrac{1}{2}\sum_n\sigma_{\varphi_n}^{(n)}\eta_{nk}\int_{0}^{t}\Omega_{n}(t')e^{i(\delta_k t+\delta\varphi_n)}dt',\end{equation}
as illustrated in Fig.~\ref{fig:displacement_and_phase}. The field amplitude $\Omega_n(t')$ and relative phase $\delta\varphi_n(t')$ controls the instantaneous amplitude and direction of displacement at time $t'$ in the phase-space of each mode.

\begin{figure}
\begin{centering}
\includegraphics[width=8.6cm]{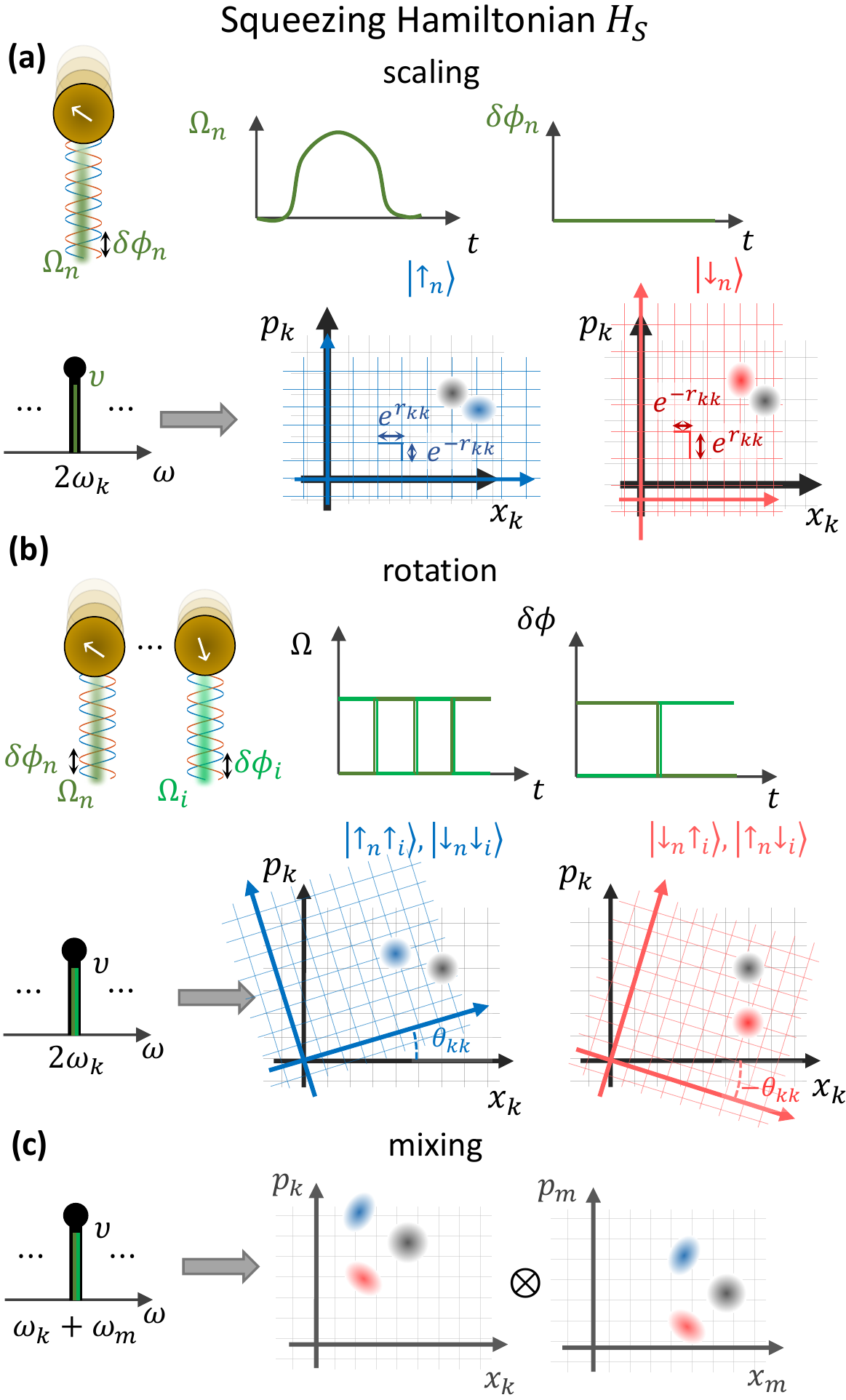}
\par\end{centering}
\centering{}\caption{\textbf{State-dependent squeezing of motion.} Tuning the frequency $\nu$ of the drive on the $n^{\textrm{th}}$ ion near twice the resonance frequencies of the vibrational modes generates a spin-dependent coordinate-transformation of phase space. (a) Spin-dependent scaling of phase-space of the $k^{\textrm{th}}$ mode. For $\nu=2\omega_k$ and $\delta \phi=0$, the squeezing axis is aligned with the $\hat{x}_k$ coordinate, resulting with dilation of $\hat{x}_k$ and contraction of $\hat{p}_k$ if the spin points upwards, and vice versa if it points downwards. (b) Spin-dependent rotation of phase-space of the $k^{\textrm{th}}$ mode. Simultaneous modulation of the phase $\delta \phi$ by even number of spins results with modulation of the squeezing axis which can rotate phase-space axes. Setting $\nu=2\omega_k$ and alternately driving two spins can generate a pure spin-dependent rotation, where the axes rotate clockwise if the two spins are aligned and counter-clockwise if the spins are anti-aligned. (c) Spin-dependent mixing of modes. Driving spins synchronously at $\nu=\omega_k+\omega_m$, the sum of resonance frequencies of two different modes $k,m$ correlate their phase-space coordinates. This correlation is manifested as two-mode squeezing or correlated rotations. Off-resonance driving exerts the interactions in (a)-(c) simultaneously. The spin-dependent scaling matrix $\boldsymbol{r}$ and rotation matrix $\boldsymbol{\theta}$ are uniquely determined by the complex transformation parameters $\psi$ and $\chi$ which are presented in the main text; see Eqs.~(\ref{eq:r_squeez}-\ref{eq:theta_squeez}). 
\label{fig:squeezing_rotation}}
\end{figure}

Tuning near the second motional sidebands with detuning $2\Delta_{k} \equiv \nu-2\omega_k$ generates the interaction Hamiltonian under the rotating wave approximation
\begin{equation}
H_{S}=\frac{\hbar}{4}\sum_{l,m,n}\eta_{nl}\eta_{nm}\Omega_{n}(t)e^{i[(\Delta_{l}+\Delta_{m})t+\delta\phi_{n}]}\sigma_{\phi_n}^{(n)}\hat{a}_{l}\hat{a}_{m}+\textrm{h.c.}\label{eq:displacement_hamiltonian_H_S}
\end{equation}
This Hamiltonian acts to instantaneously squeeze phase space coordinates as shown in Fig.~\ref{fig:squeezing_rotation}a. Here the relative motional phase between the two tones $\delta\phi_n=(\phi_{n+}-\phi_{n-})/2$ determines the axis in which phase-space coordinates are instantaneously squeezed, whereas the common phase $\bar{\phi}_n=\pi+(\phi_{n+}+\phi_{n-})/2$ specifies the projection of the spin operator over the Bloch sphere $\sigma_{\phi_n}^{(n)}$. 

For both displacement and squeezing, we assume the common phases are fixed during the evolution and set $\bar{\varphi}_n=\bar{\phi}_n=0$ such that $\sigma_{\varphi}^{(n)}=\sigma_{\phi}^{(n)}=\sigma_{x}^{(n)}$, similar to operation of the MS gate. But we allow the motional phases $\delta\varphi_n(t)$ and $\delta\phi_n(t)$ to vary in time, allowing modulation of the directions of squeezing and displacement during the operation.

\begin{figure*}[t]
\begin{centering}
\includegraphics[width=16.5cm]{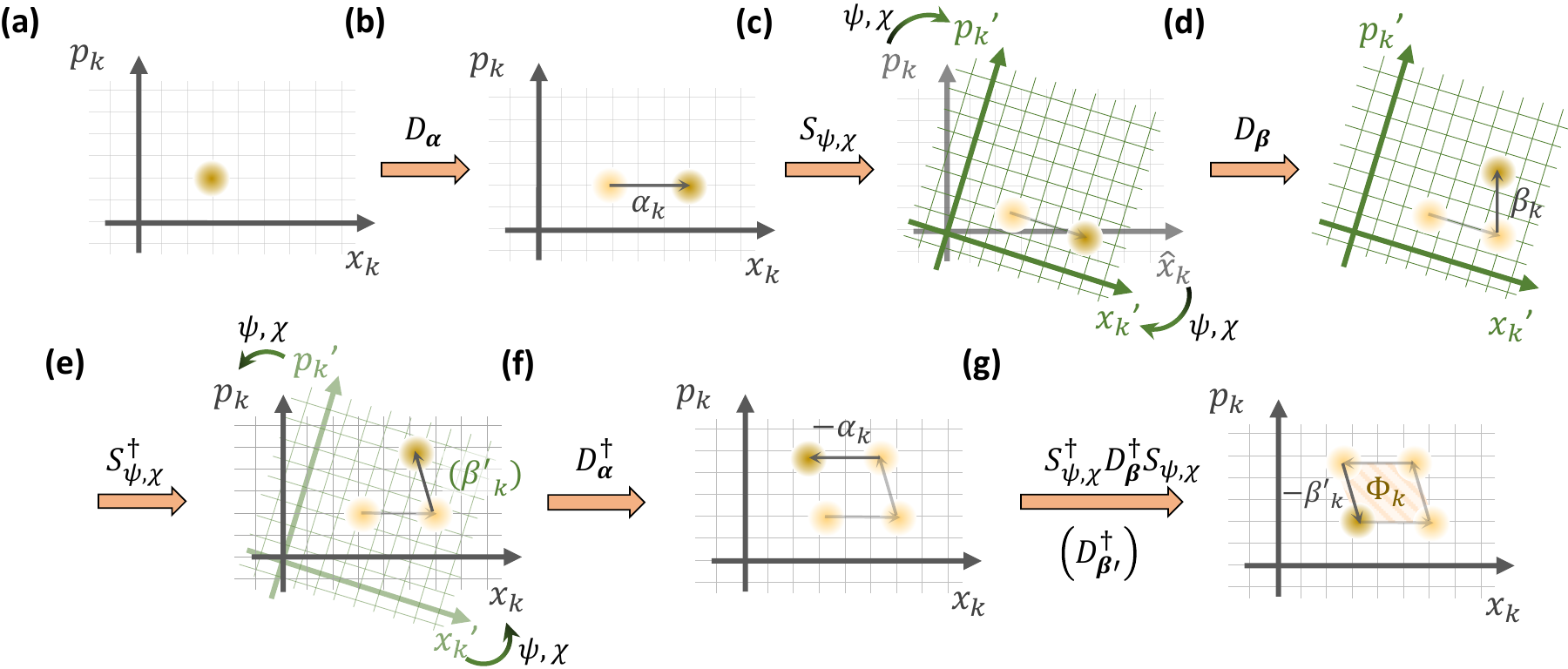}
\par\end{centering}
\centering{}\caption{\textbf{Protocol for generating high order spin-dependent Hamiltonians.} Illustration of the stages composing the unitary evolution in Eq.~(\ref{eq:lopp_operator}), constructed by alternate application of the displacement and squeezing operations. (a) The spin state of the ions and the motional state of the crystal is initially decoupled. (b) Displacing the motion of the $k^{\textrm{th}}$ vibrational mode by amount $\alpha_k$. (c) Coordinate-transformation of the phase-space with transformation matrices $\psi,\chi$ via scaling and rotations generated by the squeezing operation. The motional state remains stationary with respect to the transformed coordinates. (d) Displacement of the ions along the \textit{original coordinate frame} by amount $\beta_k$. (e) Reversing the coordinate transformation to the original frame in stage (c). The overall evolution in stages (c-e) is equivalent to net displacement by an amount $\beta'_k$ along the transformed reference frame. Importantly, as the coordinate transformation is spin-dependent, the transformed displacement $\beta'_k$ can be comprised by products of multiple spin operators.  (f) Reversal of the displacement in (b). (g) Displacement by amount $-\beta'_k$ via repeating the sequence in (c-e) but reversing the evolution in (d). Any entanglement between the spins and motion is erased, but the spins accumulate a geometric phase $\Phi_k$ that is proportional to the area enclosed in phase-space. As $\alpha_k,\beta_k'$ are spin-dependent, so also is $\Phi_k$, which corresponds to the effective spin Hamiltonian that is realized by this evolution [c.f.~Eq.(\ref{eq:effective_Hamiltonian})].
\label{fig:loop_operator}}
\end{figure*}

\section{Evolution by the Squeezing Hamiltonian}\label{sec:Spin-Motion Evolution}

The squeezing Hamiltonian in Eq.~(\ref{eq:displacement_hamiltonian_H_S}) contains quadratic motional operators with time-dependent coefficients, generating an infinite series of commutators in the evolution operator. 
Thus we solve for time evolution of the motional operators in the Heisenberg picture. Since the squeezing Hamiltonian is quadratic in the motional operators, the Heisenberg equations of motion are linear in the same operators and the time evolution can be described by the time-dependent Bogoliubov transformation of the $k^{\textrm{th}}$ phonon mode (see Appendix \ref{sec:phase-space-dynamics}):
\begin{equation}\label{eq:Bogo_trans}
\hat{a}'_k(t)=S_{\psi,\chi}^{\dagger}(t)\hat{a}_kS_{\psi,\chi}(t)=\sum_{j=1}^{M}\psi_{kj}(t)\hat{a}_j+{\chi}_{kj}(t)\hat{a}^\dagger_j
\end{equation}
where $S_{\psi,\chi}(t)$ is the unitary evolution operator generated by the squeezing Hamiltonian $H_S$. The transformation in Eq.~(\ref{eq:Bogo_trans}) is a complex symplectic \cite{cariolaro2017hamiltonians} and preserves the commutation relations of $\hat{a}_{k}$ and $\hat{a}^{\dagger}_{k}$. It therefore acts to scale, rotate and mix the phase-space coordinates of the different modes, as we illustrate in Fig.~\ref{fig:squeezing_rotation} and show in Appendix \ref{sec:phase_space_appendix}.

The transformation is determined by the complex-valued spin-dependent matrices $\psi_{kj}(t)$ and ${\chi}_{kj}(t)$.
To absorb the detuning $\Delta_k$ of the applied squeezing drives from Eq. (\ref{eq:displacement_hamiltonian_H_S}), we describe their evolution in a frame that rotates at frequency $\Delta_k$ by defining $\tilde{\psi}_{kj}=e^{i\Delta_k t}\psi_{kj}$ and $\tilde{\chi}_{kj}^{*}=e^{-i\Delta_k t}\chi_{kj}^{*}$. These rotated mixing parameters satisfy
\begin{align}
\label{eq:psi_tilde_kj_dynamics}
\partial_{t}\tilde{\psi}_{kj}=&i\Delta_k \tilde{\psi}_{kj}-\tfrac{i}{2}\sum_{m,n}\eta_{nk}\eta_{nm}\Omega_{n}e^{-i\delta\phi_{n}}\sigma_{x}^{(n)}\tilde{\chi}_{mj}^{*}\\\label{eq:xi_tilde_kj_dynamics}\partial_{t}\tilde{\chi}_{kj}^*=&-i\Delta_k \tilde{\chi}_{kj}^*+\tfrac{i}{2}\sum_{m,n}\eta_{nk}\eta_{nm}\Omega_{n}e^{i\delta\phi_{n}}\sigma_{x}^{(n)}\tilde{\psi}_{mj}
\end{align}  

Importantly, the transformation parameters $\psi_{kj}(t)$ and ${\chi}_{kj}(t)$ depend on the many body spin-state of the $N \le M$ ions that are illuminated by the driving fields, rendering an effective $N$-body interaction.
For each of the $2^{N}$ configurations of these spin states, Eqs.~(\ref{eq:psi_tilde_kj_dynamics}-\ref{eq:xi_tilde_kj_dynamics}) represent a set of $N$ linear differential equations with initial conditions $\psi_{jk}=\delta_{jk}$ and $\chi_{jk}=0$. While the total number of equations scales exponentially in the order of interaction $N$, it scales only linearly with the length of the chain $M$. The mixing transformation matrices $\psi(t)$ and $\chi(t)$ determine the unitary evolution operator $S_{{\psi},{\chi}}$ uniquely, as we show in Appendix \ref{sec:phase_space_appendix}. Control over the transformation parameters is thus sufficient to describe the quantum evolution during the squeezing operation. 

It is intriguing that the state-dependent transformation that is realized by the squeezing Hamiltonian produces operations, such as state-dependent rotations of phase space, that do not directly appear in the Hamiltonian $H_S$ in Eq.~(\ref{eq:displacement_hamiltonian_H_S}). In fact, the set of effective operations that can be realized belong to the Lie-algebra that is generated by this Hamiltonian, which we derive in Appendix \ref{sec:Lie_algebra}. We find that the group of effective Hamiltonians that can be realized corresponds to the simple symplectic Lie group $\textrm{Sp}(2M,\textbf{R})$ \cite{LieBook} for the motional operators, multiplied by products of spin operators up to $N$ order. This result is a nontrivial extension of the single mode case in Refs.~\cite{Wang2001,Katz2022Nbody}.

\section{N-body gate protocol}\label{sec:loop_protocol}

We aim to realize a unitary evolution operator, which after some time $T$ corresponds to the action of an effective spin Hamiltonian manifesting high-order interactions. As the motional state is prone to heating, dephasing and initialization errors, high fidelity manipulation of the spins usually requires the evolution to be insensitive to the initial motional state, as well as the erasure of correlations that are developed between spins and motion during the evolution. This goal underlines a two-fold challenge: engineer useful spin-dependent interactions on one hand and disentangle the states of motion and spins on the other hand. 

We focus our analysis on a simple protocol that ensures disentanglement of spins and motion at the end of the gate and generates high-order spin interactions independent of the motional state. The protocol relies on sequential and interleaved applications of squeezing and displacement operations. Independent of the number of ions in the chain $M$, or the number of target interacting spin-bodies $N$, we decompose the spin-motion evolution into the following eight stages
\begin{equation}\label{eq:lopp_operator}
U(T)= S_{\psi,\chi}^{\dagger}D^\dagger_{\boldsymbol{\beta}}S_{\psi,\chi}D^\dagger_{\boldsymbol{\alpha}}S_{\psi,\chi}^{\dagger}D_{\boldsymbol{\beta}}S_{\psi,\chi}D_{\boldsymbol{\alpha}}.\end{equation} 
Here, $D_{\boldsymbol{\alpha}}$ and $D_{\boldsymbol{\beta}}$ correspond to the displacement evolution operator generated solely by $H_D$ (Eq.~\ref{eq:displacement_hamiltonian_H_D}).
We use a vector form of displacement arguments $\boldsymbol{\alpha}$ and $\boldsymbol{\beta}$ to compactly denote the target spin-dependent displacements of all modes, with mode $k$ displaced by the expression in Eq.~(\ref{eq:displacemt_control}).
The term $S_{\psi,\chi}$ in Eq.~(\ref{eq:lopp_operator}) corresponds to the squeezing evolution operator generated solely by $H_S$ (c.f.~Eq.~\ref{eq:displacement_hamiltonian_H_S})), with $\psi$ and $\chi$ representing the target transformation matrices.

The evolution in Eq.~(\ref{eq:lopp_operator}) has a simple physical interpretation illustrated graphically in Fig.~\ref{fig:loop_operator}. Absent squeezing operations (i.e.~$H_S=0$ and $S_{\psi,\chi}=\mathbb{1}$), the ions' motion is described by closed contours in phase space of the motional-modes, leading to accumulation of a geometric phase $\Phi_k=2\textrm{Im}(\alpha_k^*\beta_k)$ by the $k^{\textrm{th}}$ phonon mode, and thus to a total accumulation of geometric phase $\Phi=\sum_k\Phi_k$ during the evolution. The spin-dependence of this phase gives rise to a quadratic spin Hamiltonian known from the usual MS operation. However, when the squeezing operation $S_{\psi,\chi}$ is interspersed, it rotates and scales phase-space coordinates while $S^{\dagger}_{\psi,\chi}$ inverts the same. Using Eq.~(\ref{eq:Bogo_trans}) we find that these two squeezing operations generate \begin{equation}\label{eq:beta_tilde} S^\dagger_{\psi,\chi}D_{\boldsymbol{\beta}}S_{\psi,\chi}=D_{\boldsymbol{{\beta'}}}\,\,\textrm{with}\,\,{\beta}_i'=\sum_k(\psi_{ki}^*\beta_k-{\chi}_{ki}\beta_k^{*}).\end{equation}  Notably, the resulting evolution has no quadratic terms but rather is linear in the motional operators. The emergent displacement vector $\boldsymbol{{\beta}'}$ corresponds to displacement by amount $\boldsymbol{\beta}$ but in a phase-space whose coordinates are scaled or rotated by the transformation matrices $\psi$ and $\chi$. Crucially, the spin-dependence of $\psi$ and $\chi$ renders $\beta'$ nonlinear in the spin operators. Therefore, the overall evolution operator $U$ in Eq.~(\ref{eq:lopp_operator}) is equivalent to the series of displacements $\boldsymbol{\alpha}\rightarrow \boldsymbol{{\beta}'}\rightarrow -\boldsymbol{\alpha} \rightarrow -\boldsymbol{{\beta}'}$ that closes all motion modes in phase space 
and results in an evolution 
\begin{equation}\label{eq:loop_phase_gate} U(T)=e^{-i\Phi}\,\,\,\textrm{with}\,\,\,\Phi=2\textrm{Im}\Bigl(\sum_i\alpha^*_i{\beta'}_i\Bigr).
\end{equation}

Here, the net spin-dependent geometric phase $\Phi$ is equivalent to the effective spin Hamiltonian \begin{equation}\label{eq:effective_Hamiltonian}H_{\textrm{eff}}=\hbar\Phi/T,
\end{equation} indices
where $T$ is the total duration of the evolution in Eq. \ref{eq:lopp_operator}.
The inherent $N$-body nature of the evolution operator and Hamiltonian appears in the spin dependences of $\alpha_i$, $\beta_i$, and in particular the mode-mixing transformation parameters $\psi_{ki}$ and $\chi_{ki}$.

\begin{figure*}[t]
\begin{centering}
\includegraphics[width=17.5cm]{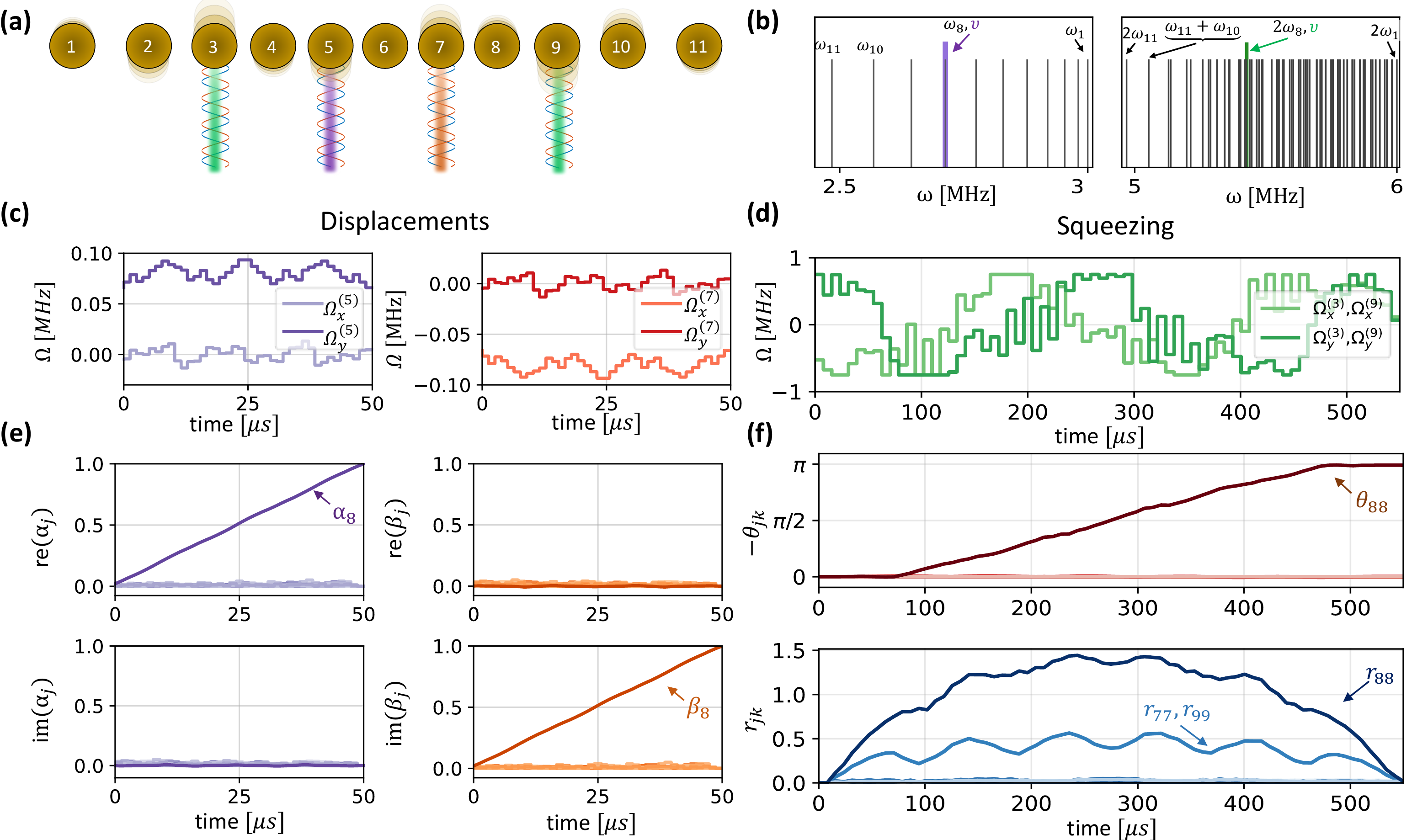}
\par\end{centering}
\centering{}\caption{\textbf{Numerical implementation of the $4$-body Stabilizer Hamiltonian} \textbf{(a)} Realization of the $4$-body stabilizer Hamiltonian $H_{\textrm{eff}}=\sigma_x^{(3)}\sigma_x^{(9)}\sigma_x^{(5)}\sigma_x^{(7)}$ in a chain of $11$ ions based on the protocol in Section \ref{subsec:stabilizer_operator} and Eq.~(\ref{eq:lopp_operator}). We use motional mode $p=8$ to generate the interaction, employing ion $5$ ($7$) to displace the position (momentum) of this mode by $\alpha_{8}$ ($\beta_{8}$), and employing ions $3,9$ to generate pure spin-dependent rotation of phase-space via squeezing. (b) Frequency tuning of the displacement and squeezing beams with respect to the first and second sideband transitions respectively. We use $\nu=\omega_8$ to generate displacements and $\nu=2\omega_8$ to generate squeezing, maximizing the coupling with the desired $p=8$ mode. Note that the ions interact with bichromatic fields which symmetrically drive both red and blue side-bands transitions, but for brevity only the blue side bands are shown. (c) Calculated phase ($\Omega_x^{(n)}(t)$) and quadrature ($\Omega_y^{(n)}(t)$) components of the control fields which generates a single displacement over $\tau_d=50\,\mu\textrm{s}$ by acting on ions $n=5,7$. 
(d) Calculated pulse shape (phase and quadrature) of the control fields simultaneously applied to ions $n=3,9$ which generate rotations over $\tau_s=550\,\mu\textrm{s}$. We consider simultaneous modulation of $\Omega_x^{(n)}$ and $\Omega_y^{(n)}$ for $n=3,9$. (e-f) Target displacements and scaling parameters for the spin state $\mid\uparrow_x^{(3)}\uparrow_x^{(5)}\uparrow_x^{(7)}\uparrow_x^{(9)}\rangle$. (e) Target displacement of $\textrm{re}(\alpha_8)=1$ along ${x}_8$ coordinate and $\textrm{im}(\beta_8)=1$ along ${p}_8$ coordinate are realized, while displacements of all other modes are erased by the end of the pulse. (f) At the end of the sequence, target rotation of $\theta_{88}=\pi$ inverts the phase-space axes of mode number $8$ while the other modes remain invariant. 
\label{fig:stabilizer_numeric}}
\end{figure*}

\begin{figure*}[t]
\begin{centering}
\includegraphics[width=17.5cm]{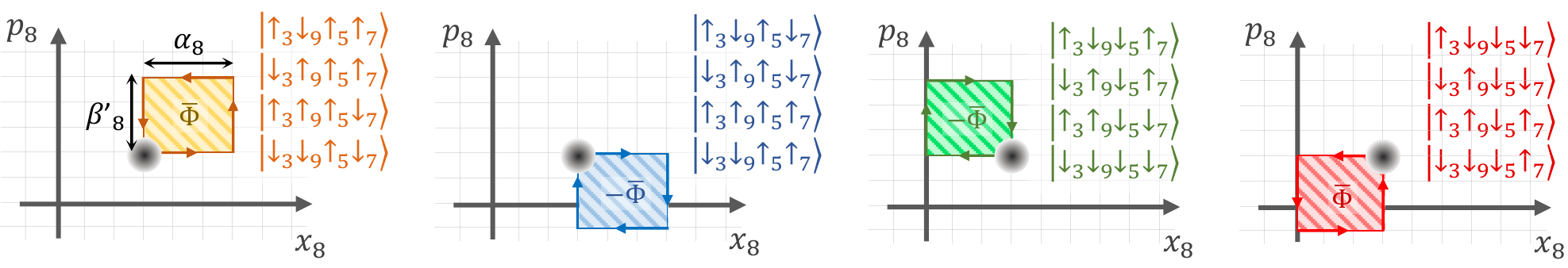}
\par\end{centering}
\centering{}\caption{\textbf{Phase-space evolution of the Stabilizer Hamiltonian}. The phase space trajectories which are used to simulate the evolution of $U=\exp(-i\bar{\Phi} \sigma_x^{(3)}\sigma_x^{(9)}\sigma_x^{(5)}\sigma_x^{(7)})$ are shown for the different $16$ spin states corresponding to the example in Fig.~{\ref{fig:stabilizer_numeric}} which uses mode number $8$. Motion is displaced rightwards by $+|\alpha_8|$ for $\mid\uparrow_5\rangle$ or leftwards by $-|\alpha_8|$ for $\mid\downarrow_5\rangle$. Similarly, motion is displaced upwards by $+\beta'_8$ for $\mid\uparrow_7\rangle$ or downwards by $-\beta'_8$ for $\mid\downarrow_7\rangle$. Here $\beta'_8=|\beta_8|\sigma_x^{(3)}\sigma_x^{(9)}$ is the displacements generated in the rotated coordinate frame conditioned on the state of ions $3$ and $9$. If ions $3$ and $9$ are aligned ($\mid\uparrow_3\uparrow_9\rangle$ or $\mid\downarrow_3\downarrow_9\rangle$) then phase space is rotated by $180^{\circ}$ yielding $\beta'_8=-|\beta_8|$, whereas for anti aligned configuration ($\mid\uparrow_3\downarrow_9\rangle$ or $\mid\downarrow_3\uparrow_9\rangle$) phase space is not rotated and $\beta'_8=+|\beta_8|$.
\label{fig:stabilizer_illustration}}
\end{figure*}

\section{Applications}\label{sec:applications}
In this section, we present two specific sequences for engineering  of particular spin-entangling gates, relying on the control over the evolution of all relevant motional modes in the chain. In subsection \ref{subsec:stabilizer_operator} we characterize application of the $N$-body stabilizer operator which is comprised of a product of $N$ spin operators, and is realized via spin-dependent rotation of phase-space. In subsection \ref{subsec:N_bit_Toffoli} we investigate a set of Hamiltonians that contain polynomials of spin operators. Finally, in subsection \ref{subsec:arbitrary_hamiltonian} we outline a systematic approach for construction of arbitrary high order spin Hamiltonians. We numerically demonstrate the control fields and characterize the performance of two examples considering a representative chain of $11$ ions in a linear Paul trap, whose parameters are detailed in Appendix \ref{sec:Optimal-control}.

\subsection{Stabilizer operator}\label{subsec:stabilizer_operator}
We consider the stabilizer operator
\begin{equation}\label{eq:stabilizer_operator} \Phi = -\bar{\Phi}\sigma_{x}^{(i_1)}\otimes\ldots\otimes\sigma_{x}^{(i_N)}\end{equation}
as the target effective Hamiltonian in Eq.~\ref{eq:effective_Hamiltonian}, for an even integer $N\leq M$, amplitude $\bar{\Phi}$ and a choice of the interacting spins labeled by the vector $\textbf{i}=(i_1,\ldots,i_N)$. Notably, the operators in Eq.~(\ref{eq:stabilizer_operator}) can be transformed into other operators in the Pauli group via application of single-qubit gates preceding and succeeding the evolution.

To construct this interaction, we use an alternating sequence of displacements and squeezing from Eq.~(\ref{eq:lopp_operator}). We consider target displacements of duration $\tau_d$, each generated by sequentially illuminating single spins $i_{N-1}$ and $i_N$ to produce
\begin{equation}\label{eq:stabilizer_displacements}\alpha_j(\tau_{d})=\delta_{jp}
A\sigma_{x}^{(i_{N-1})}\,\,\textrm{and}\,\,\beta_j(\tau_{d})=i\delta_{jp}
B\sigma_{x}^{(i_{N})},
\end{equation}
where $\delta_{nm}$ is the Kronecker delta function.
Here, the net phase-space displacement of all modes is ideally zero, except for a particular mode $p$. This mode is displaced by magnitude $A$ along the $x_p$ coordinate, followed by displacement with magnitude $B$ along the $p_p$ coordinate. Importantly, as the displacements are generated by spin-dependent forces, the sign of $\alpha_p$ ($\beta_p$), and, hence the direction of displacements, depends on the spin state of the $i_{N-1}$ ($i_{N}$) ion via the spin operator in Eq.~(\ref{eq:stabilizer_displacements}).

\begin{figure}
\begin{centering}
\includegraphics[width=8.3cm]{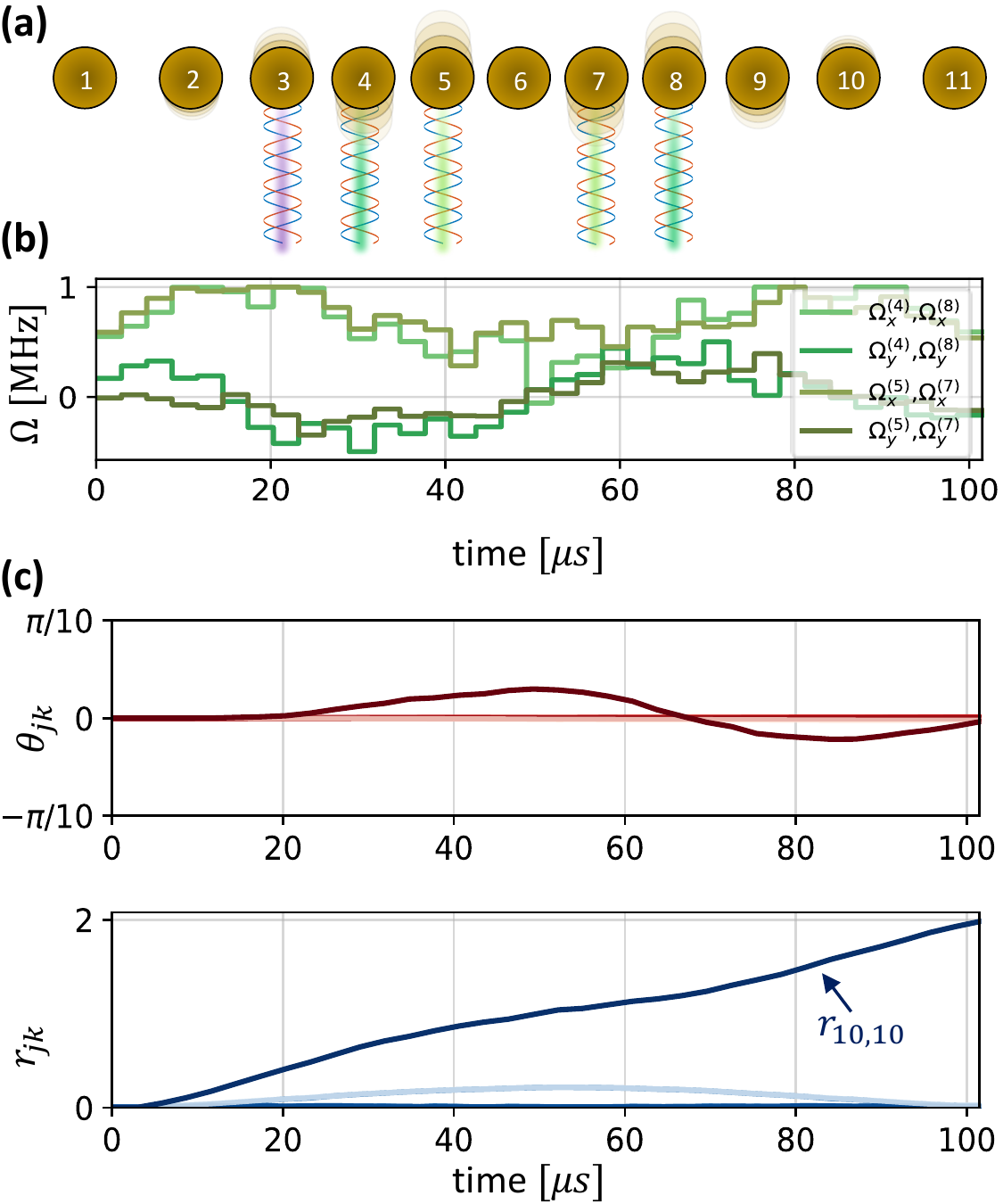}
\par\end{centering}
\centering{}\caption{\textbf{Numerical implementation of a $4$-body spin-polynomial Hamiltonian}  We exemplify the realization of a target $4$-body Hamiltonian $H_{\textrm{eff}}=(a\mathbb{1}+b\sigma_x^{(4)})(a\mathbb{1}+b\sigma_x^{(5)})(a\mathbb{1}+b\sigma_x^{(7)})(a\mathbb{1}+b\sigma_x^{(8)})$ with $a\approx1.13$ and $b\approx0.52$ in a chain of $11$ ions. 
(a) We drive ions $4,5,7,8$ with the squeezing interaction resonant with mode number $p=10$ and use ion $3$ as an auxilary ion to generate displacement in that mode. (b) Calculated pulse shape of the control fields for the squeezing interaction with $\tau_s=102\,\mu\textrm{s}$. Here the four spins are driven predominantly by $\Omega^{(n)}_x$ which squeezes the axes along the $\hat{x}_{10}$ coordinate. By symmetry of that mode, the optimal control fields for ions $n=4,8$ as well as for ions $n=5,7$ are identical. (c) Target scaling parameters for the spin state $\mid\uparrow_x^{(4)}\uparrow_x^{(5)}\uparrow_x^{(7)}\uparrow_x^{(8)}\rangle$. Target exponential scaling of $r_{10,10}=2$ squeezes $\hat{x}_{10}$ and anti-squeezes $\hat{p}_{10}$ while erasing mixing, rotations and squeezing of all other modes by the end of the pulse. The realized displacements and control fields for this case are qualitatively similar to the ones presented in Fig.~\ref{fig:stabilizer_numeric}.  
\label{fig:Toffoli_numerical}}
\end{figure}

The other $N-2$ spins participate in the desired $N$-body stabilizer Hamiltonian via the squeezing operations. We aim for a diagonal mode-mixing transformation matrix at time $\tau_s$ \cite{Wang2001,wang2002simulation}, for which \begin{equation}\label{eq:stabilizer_squeezing}
\psi_{pp}(\tau_s)=
\prod_{n=1}^{N-2}\sigma_{x}^{(i_n)} =e^{i\frac{\pi}{2}\sum_{n=1}^{N-2}\sigma_{x}^{(i_{n})}}, 
\end{equation}
and $\psi_{kk}(\tau_s)=1$ for all $k\neq p$, with $\chi_{jk}(\tau_s)=0$ for all modes.
From Eqs.~(\ref{eq:r_squeez}-\ref{eq:theta_squeez}), this transformation generates a pure  spin-dependent rotation of the phase-space of mode number $p$ by angle $\theta_{pp}(\tau_s)=\frac{\pi}{2}\sum_{n=1}^{N-2}\sigma_{x}^{(i_{n})}$ with no effect on any other mode. The phase space of mode $p$ is rotated by $180^{\circ}$ if $\frac{1}{2}\sum_{n=1}^{N-2}\sigma_{x}^{(i_{n})}$ is odd and is unaffected if it is even. 
Substitution of Eqs.~(\ref{eq:stabilizer_displacements}-\ref{eq:stabilizer_squeezing}) in Eqs.~(\ref{eq:beta_tilde}-\ref{eq:loop_phase_gate}) yields the target stabilizer operator of Eq.~(\ref{eq:stabilizer_operator}) with $\bar{\Phi}=2AB$.

We numerically simulate the operation of this gate, including the effects of all off-resonant modes of motion, with the main results shown in Fig.~\ref{fig:stabilizer_numeric}. We generate the waveforms using the optimal-control algorithm GRAPE \cite{johansson2012qutip,Pitchford2019} to search for optimal solutions of Eqs.~(\ref{eq:psi_tilde_kj_dynamics}-\ref{eq:xi_tilde_kj_dynamics}) under the target transformation parameters in Eq.~(\ref{eq:stabilizer_squeezing}), with details in Appendix \ref{sec:Optimal-control}. Here we generate the desired stabilizer interaction between $N=4$ ions in a $M=11$ long ion chain. We exemplify a gate acting on ions $\textbf{i}=(3,5,7,9)$ that is mediated predominantly by mode number $p=8$. The frequency of the beams pointing at ions $5,7$ are tuned on resonance with the first sidebands of mode $8$ ($\nu=\omega_8$; $\delta_8=0$) to generate displacement operations, setting $A=B=1$ through the control field amplitudes. The beams pointing at the other ions $3,9$ are tuned on resonance with the second sidebands of mode $8$ ($\nu=2\omega_8$; $\Delta_8=0$) to generate squeezing operations as shown in Fig.~\ref{fig:stabilizer_numeric}b. As expected, the mode spectrum of the second side-band transitions is considerably more crowded owing to nearby intermodulational sidebands between all pairs of modes. 

In Fig.~\ref{fig:stabilizer_numeric}(c-d) we present the temporal shape of the control fields using simultaneous amplitude and phase modulation for the displacement pulses $\tau_d=50 \mu s$ (c) and for the squeezing pulses (d) for $\tau_s=550 \mu s$. We express the control field $\Omega_n(t)$ on each illuminated ion in terms of its quadratures
\begin{eqnarray}\label{eq:control_field_xy}
\Omega_x^{(n)}(t)&=&\hspace{0.1in}\Omega_n(t)\sin\mu(t) \\
\Omega_y^{(n)}(t)&=&-\Omega_n(t)\cos\mu(t) \nonumber
\end{eqnarray}
with $\mu(t)=\delta\varphi(t)$ and $\mu(t)=\delta\phi(t)$ for the displacement and squeezing operations, respectively. We shape the quadrature waveforms $\Omega_x^{(n)}(t)$ and $\Omega_y^{(n)}(t)$ during displacement and squeezing stages using two different optimal-control tools, with details in Appendix \ref{sec:Optimal-control}. In Fig.~\ref{fig:stabilizer_numeric}(e-f) we present the outcome phase-space displacements and scaling parameters as a function of time for the particular case in which all spins point upwards, resulting with the target evolution. Interestingly for the squeezing evolution, both the target mode and the spectrally-nearest modes are squeezed during the pulse, yet disentangle nearly perfectly at the end of the pulse; The numerical optimization over the squeezing-operation wave-forms was terminated when the disentanglement infidelity, calculated analytically for the motional ground state and averaged over all spin configurations in the computational basis, was lower than $0.1\%$. 

In Fig.~\ref{fig:stabilizer_illustration} we illustrate the spin-dependent evolution in phase space. The spin states of ions $5$ and $7$ determine the direction of displacement along the position and momentum coordinates of mode $p=8$ in phase space respectively by setting $\alpha_8=A\sigma_x^{(5)}$ and $\beta_8=B\sigma_x^{(7)}$. Application of the target squeezing evolution rotates phase-space of the $p=8$ mode in a spin dependent manner, resulting with the modified displacement $\beta_8'=\beta_8\sigma_x^{(3)}\sigma_x^{(9)}$. Consequently, when spins $3,9$ point along the same direction in their $x$ basis, the displacement along the momentum coordinate is inverted ($\beta_8'=-\beta_8$), whereas for spins pointing at the opposite directions the displacement is unchanged ($\beta_8'=\beta_8$), therefore resulting with the geometric phase in Eq.~(\ref{eq:stabilizer_operator}). 

Interestingly, while the number operator $\hat{a}_8^{\dagger} \hat{a}_8$ that generates phase-space rotations does not appear in the Hamiltonian in Eq.~(\ref{eq:displacement_hamiltonian_H_S}), this operator is generated by sequential application of squeezing operators as discussed in Ref.~\cite{king1999quantum}. The spin-dependent rotation is generated by spin-dependent squeezing operators, such as $\hat{S}_0=\sigma_x^{(3)}({\hat{a}_{8}^{2}}-{\hat{a}_{8}^{\dagger2}})/2$ (applied when $\Omega_x^{(3)}\neq 0$) and $\hat{S}_{45}=\sigma_x^{(9)}({\hat{a}_{8}^{2}}+{\hat{a}_{8}^{\dagger2}})/2$ (applied when $\Omega_y^{(9)}\neq 0$), whose commutation yields $[\hat{S}_0,\hat{S}_{45}]=\sigma_x^{(3)}\sigma_x^{(9)}(\hat{a}_{8}^{\dagger}\hat{a}_{8}+\tfrac{1}{2})$. See Appendix \ref{sec:Lie_algebra} for further details on the set of operators that can be generated by the evolution.

\subsection{N bit spin polynomials}\label{subsec:N_bit_Toffoli}
In Ref.~\cite{Katz2022Nbody}, we proposed the uniaxial squeezing of a single motional mode to generate the target set of effective Hamiltonians  \begin{equation}\label{eq:Toffoli_operator} H = \frac{\hbar\bar{\Phi}}{T}\prod_{n=1}^{N}\left(\mathbb{1}\cosh\xi_n+\sigma_{x}^{(i_n)}\sinh\xi_n\right),\end{equation} for $N\leq M$ and positive and real $\xi_n$ where $\mathbb{1}$ denotes the identity (spin) operator. In the limit $\xi_n\gtrsim 1$ the coefficients satisfy $\cosh\xi_n\approx\sinh\xi_n\approx e^{\xi_n}/2$, and the operator in Eq.~(\ref{eq:Toffoli_operator}) becomes a projection operator which generates the N-bit controlled-phase gate, or the N-bit Toffoli gate using two additional single-qubit gates. 

Here we extend this approach and analyze the multi-mode case, which enables the squeezing of a single motional mode in a spin-dependent manner, and simultaneously erase the undesired evolution that is generated by off-resonant coupling with other modes. Here we consider the target displacements 
\begin{equation}\label{eq:Toffoli_displacements}\alpha_j=\delta_{jp}
A\sigma_{x}^{(\textrm{aux})}\,\,\textrm{and}\,\,\beta_j=i\delta_{jp}
B\sigma_{x}^{(\textrm{aux})},\end{equation} which are similar to the displacements in Eq.~(\ref{eq:stabilizer_displacements}), except here the two displacements are driven on the same auxiliary spin, which need not appear in the target Hamiltonian and can be any spin in the chain coupled to the involved modes of motion. The coupling between $N$ spins is then realized via preparing diagonal mode-mixing transformation matrices at time $\tau_s$ satisfying \begin{align}\label{eq:Toffoli_squeezing}
\psi_{pp}(\tau_s)=&\cosh\left(\sum_{n=1}^{N}\xi_n\sigma_{x}^{(i_{n})}\right)\\ {\chi}_{pp}(\tau_s)=&\label{eq:Toffoli_squeezing2}\sinh\left(\sum_{n=1}^{N}\xi_n\sigma_{x}^{(i_{n})}\right)\end{align} for a particular target mode $p$ with all other mode diagonals $\psi_{kk}(\tau_s)=1$ and $\chi_{kk}=0$. Substitution of Eqs.~(\ref{eq:Toffoli_squeezing}-\ref{eq:Toffoli_squeezing2}) in Eq.~(\ref{eq:beta_tilde}) reveals that the spin-dependence emerges via scaling the displacement along the momentum coordinate of mode $p$ by $\beta_p'=\Pi_ne^{\sigma_x^{(i_n)}\xi_n}\beta_p$, which assigns a factor $e^{\sigma_x^{(i_n)}\xi_n}$  for each spin $n$ that enlarges (compresses) the motion if the $i_n$ spin points upwards (downwards).

We demonstrate the operation of this gate in Fig.~\ref{fig:Toffoli_numerical}, generating the Hamiltonian in Eq.~(\ref{eq:Toffoli_operator}) between $N=4$ ions for an $M=11$ ion chain and for the target parameters $\xi_n=0.5$. We demonstrate the interaction between ions $(4,5,7,8)$ mediated predominantly by mode number $p=10$. The four ions are driven by beams that are tuned at $\nu=2\omega_{10}$ for $\tau_s=102\,\mu\textrm{s}$ and the control fields are presented in Fig.~\ref{fig:Toffoli_numerical}b. This pulse acts to squeeze mode $p=10$ by a factor $e^{r_{10,10}}$, and to disentangle the effect over all other modes, as presented for the case in which all the spins point upwards in Fig.~\ref{fig:Toffoli_numerical}c. For the displacements we use ion number $3$ as the auxiliary ion. 

\subsection{High-order spin Hamiltonians}\label{subsec:arbitrary_hamiltonian}
The applications in the preceding two subsections are based on spin-dependent coordinate-transformations of a \textit{single} motional mode, and the successful disentanglement of all other modes from the transformation. One strategy for generating other high-order Hamiltonians in a single step is to decompose a target spin Hamiltonian into $m\leq M$ spin-polynomials, whose structure is similar to that of Eqs.~(\ref{eq:stabilizer_operator}) and (\ref{eq:Toffoli_operator}). Then each term can be assigned to a different motional mode and the control fields for the target displacements and scaling parameters an be calculated in parallel, similar to the way in which parallel M\o{}lmer-S\o{}rensen gates are constructed \cite{lu2019global,figgatt2019parallel}.

\section{Discussion}\label{sec:discussion}
The use of spin-dependent squeezing operations between trapped atomic ion spins is a powerful technique for generating a variety of many body interactions.
By driving spin-dependent forces near the first and second side-bands, the resulting displacement and squeezing operations conspire to form families of spin-entangling gates that implement interaction between $N$ bodies, while being robust to thermal motion of the ions. We derive the Heisenberg equations of motion that enable to shape the optical fields to achieve the desired evolution over all motional modes, including those off-resonance from the targeted sidebands. Finally we numerically demonstrated and analyzed the operation of two different gate families in an eleven ion chain.

Interestingly, the spin-dependent squeezing Hamiltonian, whose terms are quadratic in the motional creation and annihilation operators, allows optical forces with a linear spin dependence to produce nonlinear spin interactions. Our representation of the squeezing action as a spin-dependent coordinate-transformation reveals the origin of this non-linearity: while the rotation angle and the squeezing parameter depend linearly on the spins, the squeezing and rotation change the motional coordinates and the underlying geometrical phase in a nonlinear manner.

Remarkably, controllable interactions can be realized despite the complex structure of the Hamiltonian. While the Magnus expansion \cite{Monroe2021,lu2019global,manovitz2017fast,martinez2021analytical} or the Wei-Norman factorization \cite{wei1963lie,sorensen2000entanglement,Katz2022Nbody} provide a description of unitary evolution under time-dependent displacement Hamiltonians, these techniques are not suitable for describing the action of the time-dependent squeezing Hamiltonian with more than one motional mode, owing to the non-terminating commutation relation of quadratic bosonic Hamiltonians. In contrast, using the time-dependent coordinate transformation in the Heisenberg picture uniquely determines the unitary evolution and, importantly, renders the control problem tractable, where the number of equations scales linearly with the number of ions in the chain. This allows the design of pulses that disentangle the spins from the motional state at the end of the gate, thus erasing any squeezing, rotations and inter-mode mixing of the motional modes that are generated during the gate. Owing to the frequency selectivity of the modes, this can be done despite of the dense mode spectrum of the second sidebands and the presence of the inter-mode coupling terms in the Hamiltonians.

This work paves the way towards efficient realization of complex building blocks for quantum computations and simulations in trapped ions systems. The tools and concepts developed in this work might also find use in other contexts. For example, it might have applications in continuous variable quantum information applications \cite{de2022error,fluhmann2019encoding,chen2021quantum,gan2020hybrid,burd2019quantum,ge2019trapped}, or in other physical systems manifesting coupling between spins and bosonic modes that act as a quantum bus, such as in superconducting circuits embedded in a microwave cavities or arrays of neutral atoms in optical cavities.

\begin{acknowledgments}
This work is supported by the ARO through the IARPA LogiQ program; the NSF STAQ program; the DOE QSA program; the AFOSR MURIs on Dissipation Engineering in Open Quantum Systems, Quantum Measurement/Verification, and Quantum Interactive Protocols; the ARO MURI on Modular Quantum Circuits; and by the U.S. Department of Energy HEP QuantISED Program through the GeoFlow Grant No. de-sc0019380.
\end{acknowledgments}

\clearpage
\appendix

 \section{Derivation of phase-space dynamics} \label{sec:phase-space-dynamics}
In this appendix, we derive Eqs.~(\ref{eq:psi_tilde_kj_dynamics}-\ref{eq:xi_tilde_kj_dynamics})  from the squeezing Hamiltonian $H_S$ in Eq.~(\ref{eq:displacement_hamiltonian_H_S}). To this end, we compute the dynamics of the motional annihilation operators $\hat{a}'_k(t)$ by the squeezing Hamiltonian in the Heisenberg picture by \begin{equation}\label{eq:Heisenberg_evolution}\partial_{t}\hat{a}'_{k}=S_{\psi,\chi}^{\dagger}\frac{i}{\hbar}\left[H_{S},\hat{a}_{k}\right]S_{\psi,\chi} \end{equation}where $\hat{a}_k$ is the time-independent annihilation operator in the interaction picture. We first calculate the commutator \begin{equation}\label{eq:commutator_tmp}\left[ H_{S},\hat{a}_{k}\right] =-\hbar\sum_{j}h_{kj}e^{-i(\Delta_k+\Delta_j)t}\hat{a}_{j}^{\dagger},\end{equation} where \begin{equation}h_{kj}=\tfrac{1}{2}\sum_{n=1}^{N}\eta_{nk}\eta_{nj}\Omega_{n}e^{-i\delta\phi_{n}}\sigma_{\phi_n}^{(n)}.\end{equation}

Application of the squeezing operators $\hat{a}'_{k}=S_{\psi,\chi}^{\dagger}\hat{a}_{k}S_{\psi,\chi}$ and substitution of the Bogoliubov transformation in Eq.~(\ref{eq:Bogo_trans}) yields
\begin{equation}\begin{aligned}\label{eq:Heisenberg_evolution3}
\partial_{t}\hat{a}'_{k}&=-i\sum_{j}h_{kj}e^{-i(\Delta_k+\Delta_j)t}\hat{a}_{j}^{\prime \dagger}\\
=&-i\sum_{j}h_{kj}e^{-i(\Delta_k+\Delta_j)t}({\chi}_{jm}^{*}\hat{a}_m+{\psi}_{jm}^{*}\hat{a}_m^{\dagger})\end{aligned}\end{equation} 
 On the other hand, direct differentiation of the the Bogoliubov transformation yields \begin{equation}\label{eq:Heisenberg_evolution4}\partial_{t}\hat{a}_k'=\sum_{m}\left(\partial_{t}\psi_{km}\hat{a}_m+\partial_{t}{\chi}_{km}\hat{a}^\dagger_m\right).\end{equation}
Comparison of Eq.~(\ref{eq:Heisenberg_evolution3}) with Eq.~(\ref{eq:Heisenberg_evolution4}) yields the dynamics of the complex-valued mode-mixing transformation matrix elements $\psi_{kj}$ and ${\chi}_{kj}$. Specifically, for a given $1\leq j,k\leq M$, calculation of the commutator $[\partial_t \hat{a}'_{k},\hat{a}^{\dagger}_{j}]$ in Eqs.~(\ref{eq:Heisenberg_evolution3}-\ref{eq:Heisenberg_evolution4}) yields

\begin{equation}\label{eq:psi_kj_dynamics}\partial_{t}\psi_{kj}=-i\sum_{m}e^{-i(\Delta_k+\Delta_m)t}h_{km}{\chi}_{mj}^{*}.\end{equation}Similarly, calculation of  $[\hat{a}_{j},\partial_{t}\hat{a}'_{k}]^*$ in Eqs.~(\ref{eq:Heisenberg_evolution3}-\ref{eq:Heisenberg_evolution4}) yields

\begin{equation}\label{eq:xi_kj_dynamics}\partial_{t}\chi_{kj}^{*}=i\sum_{m}e^{i(\Delta_k+\Delta_m)t}h_{km}^{*}{\psi}_{mj}.\end{equation} Representing the mode-mixing elments in the rotating frame ${\tilde{\psi}_{kj}=e^{i\Delta_k t}\psi_{kj}}$ and ${\tilde{\chi}_{kj}^*=e^{-i\Delta_k t}{\chi}_{kj}^*}$ in Eqs.~(\ref{eq:psi_kj_dynamics}-\ref{eq:xi_kj_dynamics}) yields Eqs.~(\ref{eq:psi_tilde_kj_dynamics}-\ref{eq:xi_tilde_kj_dynamics}).

\section{squeezing operation as coordinate transformation in phase space} \label{sec:phase_space_appendix}

We can interpret the Bogoliubov transformation in Eq.~(\ref{eq:Bogo_trans}) via the simple transformation $\hat{a}'_k={\hat{x}_k'}+i{\hat{p}_k'}=S_{\psi,\chi}^{\dagger}\hat{a}S_{\psi,\chi}$ where $\hat{x}_k'$ and $\hat{p}_k'$ are the dimensionless quadratures used to illustrate phase space in all figures. Then, the equivalent transformation of these phase-space operators in the Heisenberg picture reads
\begin{align}\label{eq:real_Bogo_trans}
\hat{x}_{k}'(t)&=\sum_{j=1}^{M}\lambda_{kj}(t)\hat{x}_{j}+\lambda_{k\tilde{j}}(t)\hat{p}_{j}\\\label{eq:real_Bogo_trans2}\hat{p}_{k}'(t)&=\sum_{j=1}^{M}\lambda_{\tilde{k}j}(t)\hat{x}_{j}+\lambda_{\tilde{k}\tilde{j}}(t)\hat{p}_{j},
\end{align}
where $\tilde{k}=k+M$ and $\tilde{j}=j+M$ for brevity. The transformation matrix $\boldsymbol{\lambda}(t)$ of size $2M\times2M$ is a function of $\psi$ and $\chi$ that is given by \begin{align}\label{eq:lambda2psi_a}\lambda_{jk}=\textrm{Re}(\psi_{jk}+\chi_{jk})\,\,&\textrm{and}\,\,\lambda_{\tilde{j}\tilde{k}}=\textrm{Re}(\psi_{jk}-\chi_{jk})\\\label{eq:lambda2psi_b} \lambda_{\tilde{j}k}=\textrm{Im}(\psi_{jk}+\chi_{jk})\,\,&\textrm{and}\,\,\lambda_{j\tilde{k}}=\textrm{Im}(\chi_{jk}-\psi_{jk})\end{align} It preserves the commutation relations of $\hat{x}_{k}$ and $\hat{p}_{k}$ at any time, and mathematically corresponds to a linear symplectic transformation. It therefore acts to scale, rotate and mix the different phase-space coordinates, as we illustrate in Fig.~\ref{fig:squeezing_rotation}. The components $\lambda_{kk},\lambda_{\tilde{k}\tilde{k}}\neq 1$ change the scaling of the $k^{\textrm{th}}$ mode's coordinates in phase-space, and the components $\lambda_{k\tilde{k}},\lambda_{\tilde{k}k}\neq 0$ correspond to their rotation. All other components correspond to correlated mixing of the different modes: $\lambda_{kj},\lambda_{\tilde{k}\tilde{j}}\neq 0$
mix the coordinates of two different modes $j\neq k$ in the form of correlated scaling (i.e.~via two-mode squeezing) and $\lambda_{k\tilde{j}},\lambda_{\tilde{k}j}\neq 0$ mix the two modes via correlated rotations (i.e.~by exchange of phonons between the modes).

We now discuss a specific representation of the coordinate transformation defined by ${\psi}$ and ${\chi}$ and show how they determine the evolution operator $S_{\psi,\chi}$. The representation we use supports the simple physical interpretation of scaling and rotation of phase space by the end of the squeezing evolution at time $\tau_s$. Based on the polar decomposition carried in Refs.~\cite{cariolaro2016bloch,cariolaro2017hamiltonians}, the complex scaling-parameters at a given time $t$ can be represented by\begin{align}\label{eq:psi_representation} \psi_{kj}(t)&=(\cosh{\boldsymbol{r}}e^{i\boldsymbol{\theta}})_{kj}\\ \label{eq:xi_representation}{\chi}_{kj}(t)&=(\sinh{\boldsymbol{r}}e^{i\boldsymbol{\vartheta}}e^{-i\boldsymbol{\theta}^T})_{kj},\end{align} where $\boldsymbol{r}(t)$,$\boldsymbol{\vartheta}(t)$ and $\boldsymbol{\theta}(t)$ are Hermitian, $M\times M$ matrices and $1 \leq j,k \leq M$. $\boldsymbol{r}(t)$ is a positive semi-definite matrix which describes the degree of squeezing of phase space at time $t$, $\boldsymbol{\vartheta}(t)$ describes the axes of squeezing in phase space at time $t$ and $\boldsymbol{\theta}(t)$ describes all phase-space rotations at time $t$. The role of these different matrices can also be seen via explicit representation of the evolution operator by \cite{ma1990multimode}
\begin{equation}\label{eq:polar_representation_U_S}S_{\psi,\chi}=e^{\frac{1}{2}\sum_{jk}\left(\hat{a}_{j}^{\dagger}\hat{a}_{k}^{\dagger}z_{jk}-\hat{a}_{j}\hat{a}_{k}z_{jk}^{\dagger}\right)}e^{i\sum_{jk}\theta_{jk}\hat{a}_{j}^{\dagger}\hat{a}_{k}},\end{equation}
where $\boldsymbol{z}(t)=\boldsymbol{r}e^{i\boldsymbol{\vartheta}}$ is a symmetric matrix which represents the multi-mode squeezing in a polar form. The first exponential map in Eq.~(\ref{eq:polar_representation_U_S}) is the multi-mode squeezing operator which mixes and scales the phase-spaces of the modes, and the second exponential term is a beam-splitter term which rotates and mixes phase space. The representation in Eq.~(\ref{eq:polar_representation_U_S}) uniquely determines the evolution operator, which establishes a relation to the transformation parameters via
Eqs.~(\ref{eq:psi_representation}-\ref{eq:xi_representation}) by \begin{align}\label{eq:r_squeez} \boldsymbol{r}=&\log\left(\sqrt{{{\psi}}{{\psi}}^{\dagger}}+\sqrt{{{\chi}}{{\chi}}^{\dagger}}\right) \\\label{eq:theta_squeez} \boldsymbol{\theta}=&-i\log\left(\left(\sqrt{{{\psi}}{{\psi}}^{\dagger}}\right)^{-1}{{\psi}}\right),\end{align}where all operations including $\log{()}$ and $\sqrt{}$ are full matrix operations. Eq.~\ref{eq:theta_squeez} is derived by inverting Eq.~(\ref{eq:psi_representation}) as $e^{i\boldsymbol{\theta}}=(\cosh\boldsymbol{r})^{-1}{\psi}$ and using $(\cosh\boldsymbol{r})^2={{\psi}}{{\psi}}^{\dagger}$. For other representations of $S_{\psi,\chi}$ see \cite{fernandez1989time,fernandez1989timeb}.

It is insightful to consider the values of the transformation $\boldsymbol{\lambda}(\tau_s)$ for some particular sets of target values. Specifically, we consider cases for which the target matrices $\boldsymbol{r}(\tau_s)$ and $\boldsymbol{\theta}(\tau_s)$ are nearly diagonal, and for $\boldsymbol{\vartheta}(\tau_s)$ that is nearly the zero matrix. The former condition minimizes the mixing between different modes by the squeezing interaction whereas the latter condition aligns the squeezing and anti-squeezing axes to be predominantly along the $\hat{x}_k$ and $\hat{p}_k$ coordinates in all $1\leq k \leq M$ phase-spaces. Under these conditions the transformation matrix is given to zeroth order by

\begin{equation}\left(\begin{array}{cc}
\lambda_{jj} & \lambda_{j\tilde{j}}\\
\lambda_{\tilde{j}j} & \lambda_{\tilde{j}\tilde{j}}
\end{array}\right)\approx\left(\begin{array}{cc}
e^{r_{jj}} & 0\\
0 & e^{-r_{jj}}
\end{array}\right)\cdot\left(\begin{array}{cc}
\cos\theta_{jj} & -\sin\theta_{jj}\\
\sin\theta_{jj} & \cos\theta_{jj}
\end{array}\right)\end{equation}
where all other $j\neq k$ coefficients are small \begin{equation}\label{eq:small_mixing}\lambda_{jk},\lambda_{j\tilde{k}},\lambda_{\tilde{j}k},\lambda_{\tilde{j}\tilde{k}}\ll 1.\end{equation}
In this representation, the phase-space coordinates transform by a two stage process. First, the $\hat{x}_j$ and $\hat{p}_j$ coordinates of the $j^{\textrm{th}}$ mode are rotated by an angle $\theta_{jj}$. Then, the rotated coordinates are scaled by a factor $e^{r_{jj}}$ along the rotated $\hat{x}_j$ and by a factor $e^{-r_{jj}}$ along the rotated $\hat{p}_j$.

\section{The reachable set of effective Hamiltonians} \label{sec:Lie_algebra}
In this appendix, we construct the Lie algebra $L$, whose elements compose the reachable set of effective time-independent Hamiltonians that can be realized by the time-dependent Hamiltonian $H_S+H_D$. This set is constructed by repeated application of the commutator operation over the operators appearing in the Hamiltonian. First, we construct the elements of the simple Lie-algebra $\textrm{span}(\mathcal{S})$ that is associated with the squeezing Hamiltonian. We do so by commuting the operators that appear only in the squeezing Hamiltonian. This Hamiltonian contains the set of operators \begin{equation}\mathcal{S}_1=\{\mathfrak{s}_{\textbf{i}}^{(1)}(\hat{a}_j^2\pm\hat{a}_j^{\dagger 2}),\mathfrak{s}_{\textbf{i}}^{(1)}(\hat{a}_j\hat{a}_k\pm\hat{a}_j^{\dagger}\hat{a}_k^{\dagger})\}\end{equation} for $1\leq j,k\leq M$ and $j\neq k$, where $\mathfrak{s}_{\textbf{i}}^{(1)}=
\sigma_{x}^{(i_1)}$ and  $1\leq i_1\leq M$. The operator $\mathfrak{s}_\textbf{i}^{(n)}$ compactly denotes a \textit{product of $n$ spin operators} by\begin{equation}\label{eq:spin_operators}{\mathfrak{s}_{i}^{(n)}=
\sigma_{x}^{(i_1)}\otimes\ldots\otimes\sigma_{x}^{(i_n)}},\end{equation} where the vector $\textbf{i}=(i_1,\ldots,i_n)$ indexes all possible spin combinations that appear in the product via $0\leq n\leq M$. Using the bosonic commutation relations $[\hat{a}_j,\hat{a}_k^{\dagger}]=\delta_{jk}$ and $[\hat{a}_j,\hat{a}_k]=0$, and the following identities \begin{equation} \left [\hat{a}_j^{2},\hat{a}_j^{\dagger 2}\right]=4\hat{a}_j^{\dagger }\hat{a}_j+\tfrac{1}{2}\,\,\textrm{and}\,\,\left[\hat{a}_{j}\hat{a}_{k},\hat{a}_{j}^{\dagger2}\right]=2\hat{a}_{j}^{\dagger}\hat{a}_{k},\end{equation} by commuting the terms in $\mathcal{S}_1$ we find the linearly independent set of operators
 \begin{equation}\mathcal{S}_2=\{\mathfrak{s}_{\textbf{i}}^{(2)}(\hat{a}_j^{\dagger }\hat{a}_j+\tfrac{1}{2}),\mathfrak{s}_{\textbf{i}}^{(2)}(\hat{a}_j^{\dagger}\hat{a}_k\pm\hat{a}_j\hat{a}_k^{\dagger})\},\end{equation}where $\mathfrak{s}_{\textbf{i}}^{(2)}=\sigma_{x}^{(i_1)}\sigma_{x}^{(i_2)}$ and  $1\leq i_1,i_2\leq M$. Commuting the different terms in $\mathcal{S}_1$ with the terms in $\mathcal{S}_2$ and using the identities \begin{equation} \left [\hat{a}_j^{2}\pm \hat{a}_j^{\dagger 2},\hat{a}_j^{\dagger }\hat{a}_j\right]=2(\hat{a}_j^{2}\mp \hat{a}_j^{\dagger 2})\,\,\textrm{and}\,\,\left[\hat{a}_{j}\hat{a}_{k}^{\dagger},\hat{a}_{j}^{\dagger 2}\right]=2\hat{a}_{j}^{\dagger}\hat{a}_{k}^{\dagger},\end{equation}
 yield the linearly independent set of operators
 \begin{equation}\mathcal{S}_3=\{\mathfrak{s}_{\textbf{i}}^{(3)}(\hat{a}_j^2\pm\hat{a}_j^{\dagger 2}),\mathfrak{s}_{\textbf{i}}^{(3)}(\hat{a}_j\hat{a}_k\pm\hat{a}_j^{\dagger}\hat{a}_k^{\dagger})\},\end{equation} where $\mathfrak{s}_{\textbf{i}}^{(3)}=
\sigma_{x}^{(i_1)}\sigma_{x}^{(i_2)}\sigma_{x}^{(i_3)}$ and  $1\leq i_1,i_2,i_3\leq M$. Similarly, commutation of the terms $\mathcal{S}_1$ with the terms in $\mathcal{S}_3$ as well as commutation of terms in $\mathcal{S}_2$ with terms in $\mathcal{S}_2$ yield the set of operators
 \begin{equation}\mathcal{S}_4\{\mathfrak{s}_{\textbf{i}}^{(4)}(\hat{a}_j^{\dagger }\hat{a}_j+\tfrac{1}{2}),\mathfrak{s}_{\textbf{i}}^{(4)}(\hat{a}_j^{\dagger}\hat{a}_k\pm\hat{a}_j\hat{a}_k^{\dagger})\},\end{equation}
 where $\mathfrak{s}_{\textbf{i}}^{(4)}=
\sigma_{x}^{(i_1)}\sigma_{x}^{(i_2)}\sigma_{x}^{(i_3)}\sigma_{x}^{(i_4)}$ and  $1\leq i_1,i_2,i_3,i_4\leq M$. It is therefore evident that for every $\mathcal{S}_n$ that is constructed by $n-1$ commutations of the terms in $S_1$, the motional operators would maintain their quadratic form and be multiplied by a product of $n$ spin operators. The set $S$ is then constructed by \begin{equation}\mathcal{S}= \mathcal{S}_1\cup \ldots \cup \mathcal{S}_M, \end{equation} which yields \begin{equation}\label{eq:Lie_algebra_squeezing}
\mathcal{S}=\{\hat{a}_k\hat{a}_m\mathfrak{s}_{\textbf{i}}^{(n_{\textrm{o}})},\hat{a}^\dagger_k\hat{a}^\dagger_m\mathfrak{s}_{\textbf{i}}^{(n_{\textrm{o}})},(\hat{a}_k^\dagger\hat{a}_m+\tfrac{1}{2}\delta_{mk})\mathfrak{s}_{\textbf{i}}^{(n_{\textrm{e}})}\},
\end{equation} where $n_\mathrm{o}$ ($n_\mathrm{e}$) run over all odd (even) values of $n$. Mathematically, if we consider the set along an eigenstate of the spin operators, then $\mathcal{S}$ corresponds to the simple Lie group $\textrm{Sp}(2M,\textbf{R})$ \cite{LieBook}.

Interestingly, $\mathcal{S}$ extends the set of operators that directly appear in the Hamiltonian $H_S$, introducing new spin-motion terms in the Unitary evolution. From the spin sector, the squeezing interaction generates products of $n>1$ spin operators, whereas $H_S$ contains a \textit{single} spin operator ($n=1$ in Eq.~\ref{eq:spin_operators}). From the motional sector, new terms that are proportional to $(\hat{a}_k^\dagger\hat{a}_m+\tfrac{1}{2}\delta_{mk})$ appear, which act to rotate the phase space coordinates as visualized in Fig.~\ref{fig:squeezing_rotation}. However, as the quadratic dependence of the motional operators in $\mathcal{S}$ is preserved, the motional identity $\mathbb{1}$ that can be associated with a motion-independent effective Hamiltonian, is not generated.

The Lie algebra associated with the total displacement and squeezing Hamiltonian is given by $L=\textrm{span}(\mathcal{S}\cup\mathcal{D})$. To construct the operators in $\mathcal{D}$, we first identify the operators that are generated solely by the displacement Hamiltonian corresponding to the sets
 \begin{equation}\mathcal{D}_1=\{\mathfrak{s}_{\textbf{i}}^{(1)}\hat{a}_j,\mathfrak{s}_{\textbf{i}}^{(1)}\hat{a}_j^{\dagger}\}\,\,\textrm{and}\,\,\mathcal{D}_2=\{\mathfrak{s}_{\textbf{i}}^{(2)}\}\end{equation} where $\mathcal{D}_1$ corresponds to the linearly independent set of operators appearing in $H_D$ and $\mathcal{D}_2$ is generated by commutation of the elements in $\mathcal{D}_1$. As the terms in $\mathcal{D}_2$ trivially commute, absent the squeezing interaction $\mathcal{D}_{\textrm{MS}}=\mathcal{D}_1\cup \mathcal{D}_2$ manifests the reachable set by the MS-type interaction, corresponding to motional displacements that are linear in the spin operators and pairwise spin-spin interactions. 
 
 With the introduction of the squeezing Hamiltonian, this reachable set can be further extended. Using the simple commutation relations $[\hat{a}_j^2,\hat{a}_j^{\dagger}]=2\hat{a}_j$ and $[\hat{a}_j^{\dagger 2},\hat{a}_j]=-2\hat{a}_j^{\dagger}$, we can commute the terms in $\mathcal{D}_1$ with the terms in $\mathcal{S}_1$ for $n\geq 1$ times and by that construct the sets $\tilde{\mathcal{D}}_n=\{\mathfrak{s}_{\textbf{i}}^{(n+1)}\hat{a}_j,\mathfrak{s}_{\textbf{i}}^{(n+1)}\hat{a}_j^\dagger\}$. Further commutation of these sets yield the motion independent set $\tilde{\mathcal{D}}_0=\{\mathfrak{s}_{\textbf{i}}^{(m)}|1 \leq m \leq M \}$. These sets can finally be united to construct \begin{equation}\mathcal{D}= \tilde{\mathcal{D}}_0\cup\tilde{\mathcal{D}}_1\cup \ldots \cup \tilde{\mathcal{D}}_{M-1}, \end{equation} which corresponds to 
  \begin{equation}\mathcal{D}=\{\hat{a}_k\mathfrak{s}_{\textbf{\textbf{i}}}^{(n)},\hat{a}_k^\dagger\mathfrak{s}_{\textbf{\textbf{i}}}^{(n)},\mathbb{1}\mathfrak{s}_{\textbf{i}}^{(n)}\},\label{eq:Lie_algebra_displacements}
\end{equation} thus containing the target $n$-body terms $\mathbb{1}\mathfrak{s}_{\textbf{i}}^{(n)}$ which we aim to generate.
 
 \section{Numerical implementation of Optimal control Solver} \label{sec:Optimal-control}
In this appendix, we describe the optimal control tools used to compute the control fields for the displacement and squeezing operations in section \ref{sec:applications}. We first describe the system parameters for which the calculation is demonstrated. 
We consider a linear chain of $11$ ions in a quadratic potential. We assume the single ion axial and secular radial frequencies $\omega_z=0.39\,\textrm{MHz}$ and $\omega_r=3\,\textrm{MHz}$ which determine the ions positions, the mode spectrum and the mode participation factors. We order the radial modes that are used for coupling the ions in a decreasing order, corresponding to the ordered set of frequencies $\omega_k\in\{3,2.981,2.954,2.919,2.878,2.830,2.775,2.713,$ $2.645,2.569,2.484\}\,\textrm{MHz}$ for $1\leq k \leq 11$. We assume a single-ion Lamb-Dicke parameter of $\eta\equiv \delta K\sqrt{\hbar/2\mathcal{M}\omega_r}=0.1$ for the driving field, and that the bichromatic field couples to the modes along a single radial axis. We also limit the drive field amplitude quadratures of each ion  $\Omega_x^{(n)}$ and $\Omega_y^{(n)}$ to be $\lesssim 1\,\textrm{MHz}$.
\subsection{optimal control of spin-dependent displacements} 
We use a simple optimal control tool to calculate the control fields $\Omega_x^{(n)}(t),\Omega_y^{(n)}(t)$ for a specific ion $n$ given a target displacements vector $\boldsymbol{\alpha}(\tau_d)$. Standard optimization tools that calculate the temporal shape of the control fields for the MS gate typically require disentanglment conditions for all modes and a target accumulated geometric phase. To realize the protocol in section \ref{sec:loop_protocol} we instead aim for a nonzero displacement vector but have no requirement on the geometric phases that are accumulated in a single stage of the evolution, owing to driving a single spin at a time. 

We assume the control fields are decomposed into $N_d$ intervals of duration $\tau$, maintaining a constant amplitude in each segment. Mathematically, they take the form $\label{amplitude_phase_modulation_pulse}\Omega_q^{(n)}(t)=\sum_{p=1}^{N_d}\Omega_{q,p}^{(n)} w(t/\tau,(p-1),p)$ for $q\in\{x,y\}$ where $w(t/\tau,(p-1),p)$ is the rectangular window function returning $1$ if $(p-1)\tau \leq t\leq p\tau$ and zero otherwise. $\Omega_{q,p}^{(n)}$ are the list of $2N_d$ amplitudes we aim to find and $\tau_d=N_d\tau$ is the overall pulse duration. We use $\tau_d=50 \mu\textrm{s}$ and $N_d=40$. 

For driving a single spin, the target complex displacements correspond to Eq.~\ref{eq:displacemt_control}, whose matrix form is given by 
\begin{widetext}\begin{equation}\label{eq:displacements_optimal_control_optim}
{\left(\begin{array}{c}
\text{Re}(\alpha_{n1})\\
\vdots\\
\text{Re}(\alpha_{nM})\\
\text{Im}(\alpha_{n1})\\
\vdots\\
\text{Im}(\alpha_{nm})
\end{array}\right)=\begin{pmatrix} \
d_{11} & \cdots & d_{1M} & \tilde{d}_{11} & \cdots & \tilde{d}_{1M}\\
\vdots & \ddots & \vdots & \vdots & \ddots & \vdots\\
d_{M1} & \cdots & d_{MM} & \tilde{d}_{M1} & \cdots & \tilde{d}_{MM}\\
\tilde{d}_{11} & \cdots & \tilde{d}_{1M} & -d_{11} & \cdots & -d_{1M}\\
\vdots & \ddots & \vdots & \vdots & \ddots & \vdots\\
\tilde{d}_{M1} & \cdots & \tilde{d}_{MM} & -d_{M1} & \cdots & -d_{MM}
\end{pmatrix} \left(\begin{array}{c}
\Omega_{x,1}^{(n)}\\
\vdots\\
\Omega_{x,N_{d}}^{(n)}\\
\Omega_{y,1}^{(n)}\\
\vdots\\
\Omega_{y,N_{d}}^{(n)}
\end{array}\right)}
\end{equation}\end{widetext}
Here we use the $2M\times2N_d$ matrix $\boldsymbol{d}$ whose elements are given by
\begin{align}\label{eq:d_elements} d_{kp}=&-\eta_{nk}\textrm{sinc}(\tfrac{\delta_k\tau}{2})\cos\left((p+\tfrac{1}{2})\delta_k\tau\right), \\\label{eq:d_elements2} \tilde{d}_{kp}=&+\eta_{nk}\textrm{sinc}(\tfrac{\delta_k\tau}{2})\sin\left((p+\tfrac{1}{2})\delta_k\tau\right),\end{align} 
for $1\leq j \leq M$ and $1\leq p \leq N_d$.
 
 For the case $N_D> M$ considered here, Eq.~(\ref{eq:displacements_optimal_control_optim}) has an infinite number of solutions, meaning that there are many phase-space trajectories that can end at the target displacements vector at time $\tau_d$. Here we calculate a single solution by applying the Moore-Penrose pseudo inversion in Eq.~(\ref{eq:displacements_optimal_control_optim}). This particular operation yields the control field vector whose norm is least among all solutions, corresponding to the waveform with least average power.
 
\subsection{optimal control of spin-dependent scaling and rotations} 
To find the control fields $\Omega_{q}^{(n)}(t),\delta\phi^{(n)}(t)$ that yield the target mode-mixing parameters $\psi_{kj}(\tau_s),\chi_{kj}(\tau_s)$ we use the open-source quantum optimal control algorithm GRAPE implemented in python \cite{johansson2012qutip,Pitchford2019}. To account for the spin-dependent dynamics, we represent the transformation parameters in the rotating frame in a compact vector format  \begin{equation}\label{eq:psi_xi_def}
\boldsymbol{\bar{\psi}}_j=(\tilde{\psi}_{1j},\ldots,\tilde{\psi}_{Mj},\tilde{\chi}_{1j}^*,\ldots,\tilde{\chi}_{Mj}^*)^T \end{equation} and explicitly account for the spin state of the driven ions using the extended basis $\boldsymbol{\psi}_j=\boldsymbol{\bar{\psi}}_j\otimes|\sigma_i\rangle$. Here $|\sigma_i\rangle$ are the computational basis vectors of the spins in the $x$ basis (i.e.~corresponding to the eigenstates of the $\sigma_{x}^{(m)}$ operators with eigenvalues $\pm1$ for all $1\leq m\leq M$) which enable the representation of all $1\leq i\leq 2^{M_S}$ spin configurations in Hilbert space. For practical implementation, we consider only the spin states that are associated with the $M_S\leq N$ ions that are driven by the squeezing beams. While the Hilbert space grows exponentially with $M_S$, we importantly note that $M_S$ scales with the order of the interaction $N$ and not with the number of ion spins in the chain $M$. For the applications we consider in this work, the exponential increase is modest because $M_S=N-2=2$ for the four body gate in section \ref{subsec:stabilizer_operator} and $M_S=N=4$ for the polynomial spin operator in section \ref{subsec:N_bit_Toffoli}. 

To render the time-dependent transformation in Eqs.~(\ref{eq:psi_tilde_kj_dynamics}-\ref{eq:xi_tilde_kj_dynamics}) compatible with the formalism of GRAPE, we cast them in the form \begin{equation}
\label{eq:psi_xi_dyn}
\partial_t\boldsymbol{\psi}_j=\mathcal{H}\boldsymbol{\psi}_j,
\end{equation} where the symmetric matrix $\mathcal{H}$ is  given by
\begin{equation}\label{eq:sympl_Hamiltonian}
\mathcal{H}=is_z\otimes\boldsymbol{\Delta}\otimes\mathbb{1}_{\sigma}+\tfrac{1}{2}\sum_{n=1}^M(\Omega_x^{(n)}s_x+\Omega_y^{(n)}s_y)\otimes\boldsymbol{\eta}_n^{2}\otimes\sigma_{x}^{(n)}.
\end{equation} We use $\boldsymbol{\eta}_n^{2}$ to denote the $M\times M$ matrix whose elements $(\boldsymbol{\eta}_n^{2})_{jk}=\eta_{nj}\eta_{nk}$ describe the coupling between the $j$th and $k$th modes via the $n$th ion. $\boldsymbol{\Delta}$ denotes a diagonal $M\times M$ matrix with nonzero elements $\Delta_k$ on the diagonal. We denote by $s_x,s_y,s_z$ the $2\times2$ Pauli matrices, which are unrelated to the spin operators, but rather construct the correct relations between the mode-mixing parameters $\tilde{\psi}_{jk}$ and $\tilde{\chi}^*_{jk}$ in Eq.~(\ref{eq:psi_xi_def}). For clarity, we denote the identity spin matrix by $\mathbb{1}_{\sigma}$.

In this form, the operator $\mathcal{H}$ is a $(2^{M_S+1}M)\times (2^{M_S+1}M)$ matrix that can be decomposed into the time-independent drift Hamiltonian $H_{\textrm{drift}}=is_z\otimes\Delta\otimes\mathbb{1}$, and the $2M_S$ control Hamiltonians taken from the set $\{s_x\otimes\boldsymbol{\eta}_{\textbf{i}_n}^{2}\otimes\sigma_{x}^{(\textbf{i}_n)},s_y\otimes\boldsymbol{\eta}_{\textbf{i}_n}^{2}\otimes\sigma_{x}^{(\textbf{i}_n)}\}$ with $1\leq n\leq 2M_S$ where the vector $\textbf{i}_n$ denotes the indices of the interacting ions. We simultaneously solve these equations by considering an optimization towards an objective ``gate'' $X(\tau_s)$ whose columns are composed of the target vectors $\boldsymbol{\psi}_j$. As the dimensions of $X$ are $2^{M_S+1}M\times 2^{M_S}M$ we technically expand it into a rectangular matrix by adding the $2^{M_S}M$ columns vectors $(\tilde{\chi}_{1j},\ldots,\tilde{\chi}_{Mj},\tilde{\psi}_{1j}^*,\ldots,\tilde{\psi}_{Mj}^*)^T\otimes|\sigma_i\rangle$ for $1\leq i\leq 2^{M_S}$ and $1\leq j\leq M$ (which physically corresponds to the transformation of $\hat{a}^{\dagger\prime}$). We also assume that $X(0)$ is the identity matrix. As the dynamics is not Unitary but rather \textit{complex} symplectic, we use the "GEN\_MAT" dynamic evolution type of the algorithm, the "trace-difference" as the fidelity measure, and the BFGS algorithm for the optimization method. For the calculation in this work we assume that the control fields are composed of up to $70$ segments.

\bibliography{Refs}

\end{document}